\documentclass[11pt,superscriptaddress]{revtex4-2}

\usepackage{graphicx}
\usepackage{dcolumn}
\usepackage{bm}
\usepackage[table]{xcolor}
\usepackage{graphicx}
\usepackage{amsmath}
\usepackage{adjustbox}
\usepackage{placeins}
\usepackage[T1]{fontenc}
\usepackage{lipsum}
\usepackage{csquotes}
\usepackage{colortbl}
\usepackage{hyperref}
\usepackage{bm}
\usepackage{multirow}
\usepackage{booktabs}
\usepackage{graphicx}
\usepackage{multirow}
\usepackage{booktabs}
\usepackage{graphicx}
\usepackage{float}
\usepackage{lineno}

\usepackage{svg}
\usepackage[normalem]{ulem}

\begin{document}


















\title{Transferable excited-state dynamics enable screening of fluorescent protein chromophores}

\author{Rhyan Barrett \textsuperscript{†}}
\affiliation{Leipzig University, Wilhelm Ostwald Institute for Physical
and Theoretical Chemistry, Linnéstraße 2, 04103 Leipzig, Germany}
\author{Sophia Wesely \textsuperscript{†}}
\affiliation{Leipzig University, Wilhelm Ostwald Institute for Physical
and Theoretical Chemistry, Linnéstraße 2, 04103 Leipzig, Germany}
\author{Julia Westermayr}
\email{julia.westermayr@uni-leipzig.de}
\affiliation{Leipzig University, Wilhelm Ostwald Institute for Physical
and Theoretical Chemistry, Linnéstraße 2, 04103 Leipzig, Germany}
\affiliation{Center for Scalable Data Analytics and Artificial Intelligence (ScaDS.AI), Dresden/Leipzig, Germany}
\affiliation{\dag These authors contributed equally}

\date{\today} 

\begin{abstract}
Transferable excited-state dynamics offer a route to efficient screening of photophysical behavior across molecular systems, but conventional nonadiabatic simulations remain prohibitively expensive. Here we introduce X-MACE, a transferable machine-learning potential for excited-state dynamics that predicts multiple potential energy surfaces, forces and oscillator strengths, and combine it with curvature-driven surface hopping to enable data-efficient screening of photochemical pathways. We apply this framework to fluorescent chromophores as an example application, using green fluorescent protein chromophore variants to demonstrate how subtle structural modifications reshape excited-state relaxation, lifetimes and photoisomerization yields. Fine-tuning a single pretrained model with fewer than 100 reference geometries per derivative yields accurate dynamics across a chemically diverse set of analogues. The screening reveals two governing design principles: steric crowding on the phenolate ring lowers the torsional barrier and accelerates access to twisted conical intersections, whereas conjugation extension stabilizes planar excited-state configurations, suppresses non-radiative decay and prolongs fluorescence. More broadly, this workflow provides a general framework for scalable excited-state screening and interpretable design of photophysical properties.

\end{abstract}

\maketitle

\section{Introduction}


The ability to visualize cellular processes using fluorescent proteins has revolutionized our understanding of life at the molecular scale. A major advance came with the discovery and development of the green fluorescent protein (GFP), recognized by the Nobel Prize in Chemistry in 2008 \cite{shimomura2022discovery}, which fluoresces upon light excitation to enable real-time tracking of dynamics in living systems. This breakthrough launched a new era of cellular imaging that has transformed disease research, diagnostics and therapies.\cite{luo2002subthalamic,herndon2002stochastic, bey2017efficient} 
Clarifying photochemical mechanisms in fluorescent proteins can thus guide rational optimization and design of fluorescent proteins, improving brightness, photostability, maturation, and spectral properties thereof. 

At the centre of GFPs fluorescence behavior is the p-hydroxybenzylidene-imidazolinone anion (HBDI$^-$), the chromophore of GFP (see Figure 1a). It is known that, upon photoexcitation, its relaxation occurs through competing radiative and non-radiative pathways, where the dominant non-radiative channel is commonly associated with twisted intramolecular charge transfer that can drive partial photoisomerisation.\cite{list2024chemical} This twisting can proceed via rotation about either the imidazolinone dihedral or the phenolate dihedral (Fig.~\ref{fig:method}a), and is coupled to pyramidalisation at the methine-bridge carbon, providing access to efficient $S_1/S_0$ decay.\cite{list2024chemical,list2022internal} Consistent with this picture, HBDI$^-$ exhibits a pronounced temperature dependence: fluorescence is enhanced at low temperature and suppressed as temperature increases, indicating increased access to non-radiative relaxation channels at higher internal energies.\cite{martin2004origin, svendsen2017origin} This behaviour suggests that the $S_1/S_0$ conical intersection seam is not readily reachable from the equilibrium excited-state minimum, but instead requires substantial structural distortion, specifically large-amplitude torsion about either ring toward roughly 90 degrees, to reach torsional conical intersections as identified in prior electronic structure analyses.\cite{list2022internal,martin2004origin} Mechanistically, the excitation from non-bonding to the anti-bonding orbital permits bridge torsion, while sufficient kinetic energy is needed to attain the distorted geometries, directly linking non-radiative yield to temperature; accordingly, improving GFP performance can be framed as shifting the balance toward radiative decay by restricting these torsional deactivation routes, thereby enhancing fluorescence efficiency for bioimaging applications.

A great advantage of fluorescent proteins such as GFP is the existence of a diverse family of variants, whose excited-state properties can be tuned by subtle structural changes to their chromophore and surrounding protein environments.\cite{meech2009excited, nienhaus2016chromophore} Consequently, many studies focus on the effects of small variations in the chromophore on fluorescent properties.\cite{conyard2011chemically, ashworth2023alkylated} However, current studies are limited to a few individual molecules due to the time-consuming nature of experiments and design processes. Theoretical modelling could overcome this challenge by screening many chromophores prior to any wet lab work, but this process demands accurate theoretical non-adiabatic excited-state dynamics simulations that are prohibitively expensive: even for a single HBDI$^-$ derivative, the required sampling of different conformations to achieve statistically significant dynamics can consume years of compute time, and extending this to a substituted series of chromophores rapidly scales to hundreds of years on high-performance resources. Although recently emerged machine learning potentials can alleviate these costs, most existing excited-state approaches still require large, molecule-specific training datasets for each chromophore to be simulated, making systematic screening across diverse HBDI$^-$ variants difficult in practice.

Foundational models offer a promising route toward screening, as they can learn from large, heterogeneous datasets and transfer this knowledge to new systems with little to no additional data. While such models have already transformed ground-state simulations, they have not yet been applied to photodynamics screening. Here, we introduce the first step towards a foundational model for excited state dynamics and a general framework for screening excited-state dynamics of hundreds of chromophores. Applied to a diverse set of HBDI$^-$ derivatives, the approach enables efficient comparative simulations and provides mechanistic insights into key factors governing their dynamics. By systematically simulating many variants, we show that increased conjugation leads to substantially longer lifetimes and sustained $S_1$ population, resulting in a larger fraction of radiative de-excitation and consequently, enhanced fluorescence potency of the HBDI$^-$ system.

\section{Results and Discussion}

\subsection{Machine Learning Workflow}

\begin{figure*}[htbp]
\centering
\includegraphics[width=\textwidth]{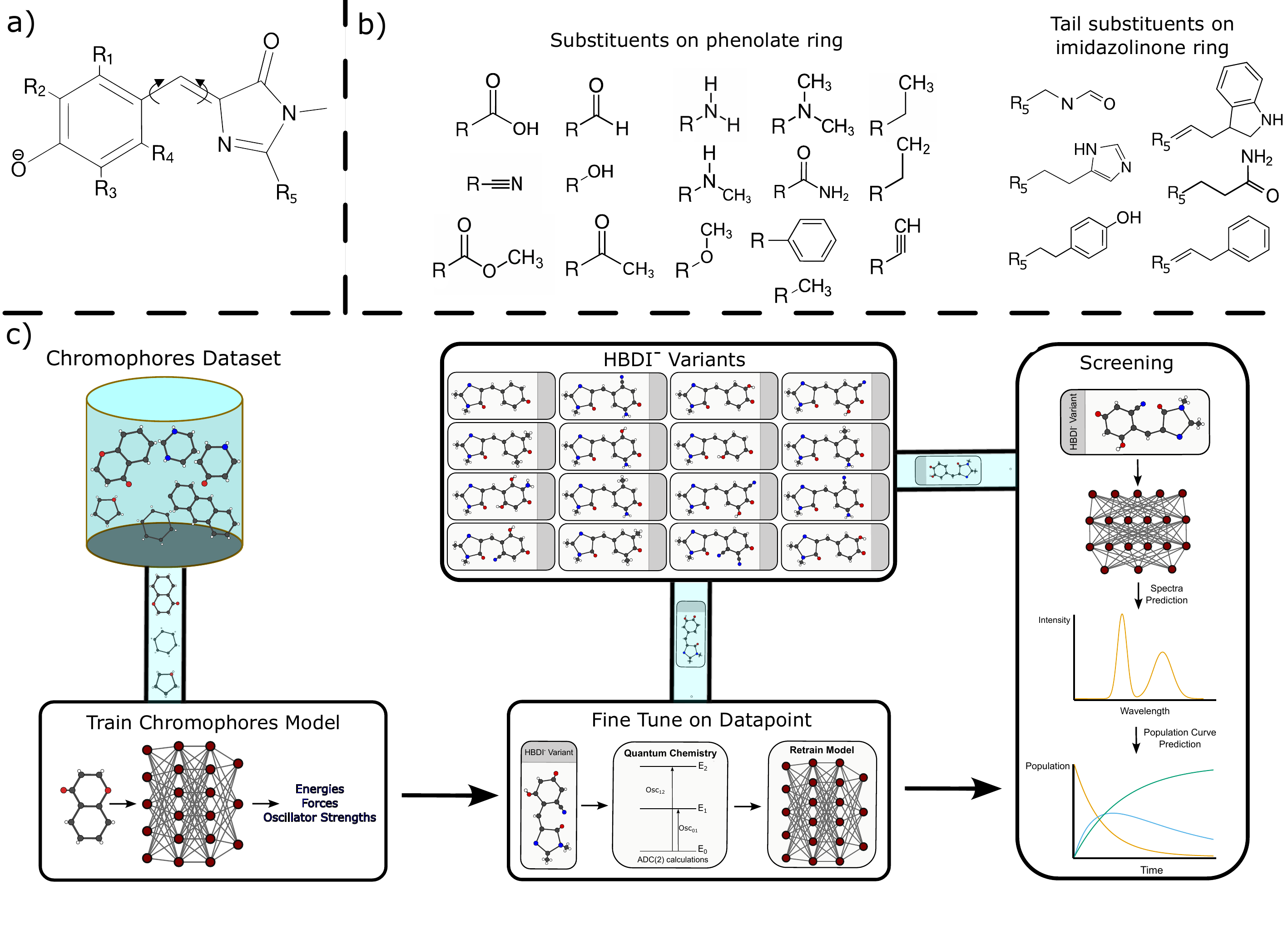}
\caption{(a) Diagram of the HBDI$^-$ molecule, highlighting the possible sites for structural modification. (b) Representative functional groups that were attached to the molecule. Electron-donating and electron-withdrawing substituents are randomly assigned to one of the positions between $R_1$ and $R_4$, while bulkier tail substituents placed at position $R_5$. (c) Schematic overview of the workflow used throughout this study. An initial model is trained on the energies, forces, and oscillator strengths of a diverse set of chromophores. This pretrained model is then fine-tuned on a small number of geometries for each HBDI$^-$ variant. The resulting model is subsequently employed to predict the absorption spectra and population dynamics of all variants.}
\label{fig:method}
\end{figure*}

To efficiently assess how structural changes influence photophysical and photochemical behaviour across a range of chemical systems, a two-stage machine learning workflow for excited-state modelling and screening (Fig. \ref{fig:method}c) was developed. As illustrated, the first step requires an excited-state surrogate model that needs to be trained on a chemically diverse collection of chromophores to learn general relationships between molecular structure and excited-state properties. Therefore, we extend the message-passing atomic cluster expansion model (MACE)\cite{batatia2022mace}, which has previously shown high generalizability between systems, to multiple excited state potentials, forces, and oscillator strengths, and call it X-MACE. For training, we first compile a dataset of ~12,000 chromophores with details on the dataset and quantum-chemical calculations provided in the methods section.
The resulting X-MACE model is then adapted to a specific chromophore system using fine-tuning on a small reference dataset that is, ideally, representative of the relevant conformational space of each chromophore. In this work, we fine tune the model on a family of HBDI$^-$ derivatives with systematic substitutions that are illustrated in Fig. \ref{fig:method}b. These derivatives are generated by combining single substitutions at different positions on the phenolate ring of HBDI$^-$ (panel a) with substitutions on the imidazolinone ring, yielding a broad set of variants. To construct the fine-tuning dataset, we sample geometries for each HBDI$^-$ derivative along the relevant torsional and pyramidalisation coordinates and generate additional geometries via Wigner sampling. This step specializes the model to enable accurate predictions for the target system while retaining the transferable knowledge learned during pretraining (Fig.\ref{fig:method}c middle). An amount of approximately 50-100 data points per system is found to be sufficient, while earlier excited-state machine learning models usually required several hundred thousand of data points for nonadiabatic molecular dynamics.\cite{axelrod2022natcomm,westermayr2022deep} 
After fine-tuning, the model is used as a fast screening engine: predicted oscillator strengths can be used to generate absorption spectra and define initial conditions for excited-state dynamics simulations, while the learned excited-state potential energy surfaces and corresponding forces enable nonadiabatic molecular dynamics (see Fig. \ref{fig:method}c right). For photodynamics, X-MACE is interfaced with the SHARC program for surface hopping molecular dynamics.\cite{mai2018nonadiabatic, mai2025sharc} Details on the dynamics settings can be found in the methods section. In total, ~19,300 trajectories were run for a number of 193 chromophores. Throughout the dynamics, X-MACE retains accuracy across chemically diverse HBDI$^-$ derivatives even in the non-equilibrium, highly distorted geometries where ML transferability typically fails. Therein highlighting, how adapting a single pretrained model with fewer than 100 reference geometries per variant preserves key dynamical observables (population evolution, lifetimes, photoisomerisation yields) and enables excited-state screening to scale from static single-point calculations to statistically robust trajectory ensembles.

\begin{figure*}[htbp]
\centering
\includegraphics[width=\textwidth]{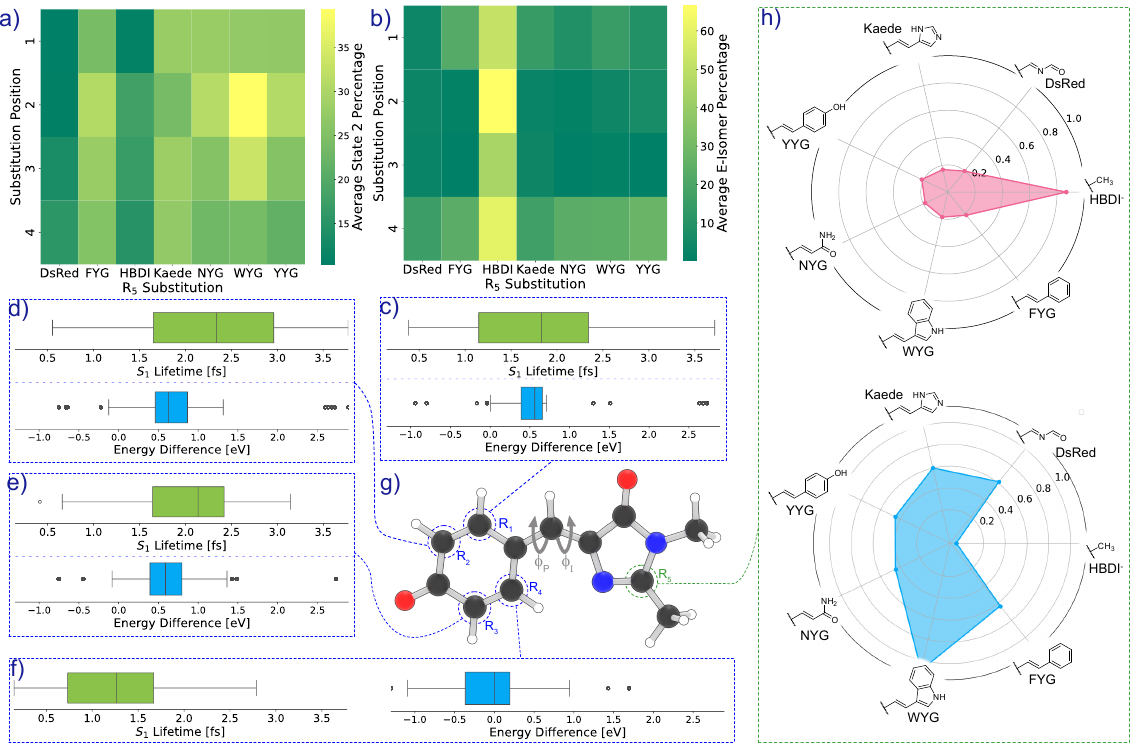}
\caption{Overview of substitution-dependent effects on excited-state properties of HBDI$^-$ derivatives. Panels a and b show heat maps of combined phenolate- and imidazolinone-substitution patterns and their effects on lifetime-related and isomerisation-related observables. The distributions of lifetime values and $S_1$ energy differences between planar and twisted structures ($\phi_I\approx 90°$) for derivatives grouped by substitutions at phenolate positions are shown in: c) $R_1$, d) $R_2$, e) $R_3$ and f) $R_4$. g) The molecular scaffold and the substitution sites analysed in this work, including $R_1 - R_4$ and $R_5$ position labels and torsion indication. Panel h) summarizes$R_5$ substitution effects, with radar plots showing the average HBDI$^-$ backbone pseudo-partial charge (top) and the energy difference between planar and twisted imidazolinone structures (bottom) for different $R_5$ substituents.}
\label{fig:overview}
\end{figure*}

\subsection{Dynamics Screening}

To illustrate the relationships between molecular structure and photochemical properties in more detail, we first group the screened HBDI$^-$ derivatives according to the substitution pattern on the phenolate ring (positions R$_1$–R$_4$, Fig.~\ref{fig:method}a). 
Quantification of how chemical substitutions at the phenolate ring modification sites reshape the relaxation behaviour of HBDI$^-$, was achieved by extracting two primary observables from the trajectory ensemble, which are the excited-state (S$_1$) population percentage at the end of the simulation time of 7~ps and the photoisomerisation percentage from the Z to the E isomer. 

Fig.\ref{fig:overview}a) and b) summarize these findings. The overall trend shows that, as expected, variants exhibiting longer $S_1$ lifetime display lower isomerisation yield. This inverse relationship follows directly from the main requirement for isomerisation, in particular that productive isomerisation proceeds through non-radiative decay at the $S_1/S_0$ conical intersection, which is accessed only after the chromophore reaches a strongly twisted geometry between the two rings along the bridge (Fig.\ref{fig:method}a)). Systems with increased lifetimes are therefore those that, on average, either encounter a larger torsional barrier or otherwise sample the conical intersection region less efficiently on the simulation timescale, reducing the probability of crossing to $S_0$ in the twisted configuration and suppressing isomerisation.

To further analyse the influence of substitutions, the mean excited-state lifetimes with the corresponding torsional activation barriers are compared for each group (Fig.~\ref{fig:overview}c–f). These plots show the mean and the distribution of the S$_1$ lifetimes for each system class and the $S_1$  energy difference between the planar and $\phi_I \approx 90°$ structures.
Focusing on steric effects, we observe a clear trend: substitution at $R_4$ has the strongest negative impact on the prolongation of excited-state lifetimes (panel f). Due to steric repulsion with the adjacent imidazolinone ring, $R_4$ substitution lowers the effective torsional barrier, facilitating access to twisted geometries associated with decay through the conical intersection. As a result, $R_4$-substituted derivatives exhibit the shortest lifetimes of approximately 1.3 ps and near-zero activation barriers. Comparably, substitution at $R_1$ (panel c) introduces a similar but weaker steric repulsion, leading to a smaller barrier decrease and correspondingly more modest changes in the dynamics: on average, $R_1$ variants show a slightly reduced lifetime of about 1.8 ps. In contrast, substitutions at $R_2$ and $R_3$ (panels d and e, respectively) introduce minimal steric influence along the torsional coordinate. Consequently, these variants show little systematic change in their dynamics, retaining longer lifetimes and higher activation barriers.

To systematically extend the conjugation length, a $\pi$-connected tail at position $R_5$ (Fig.~\ref{fig:method}a) was introduced. We found that this additional conjugation modifies the electronic density redistribution upon excitation, see panel h in Figure \ref{fig:overview}. The plots show the different substituents at position R$_5$. The top panel (reddish) shows the pseudo-partial charges distributed only at the HBDI$^-$ backbone, while the lower panel (blueish) shows the energy difference between the planar and the $\phi_I$ twisted structures. To quantify substituent-dependent changes in charge redistribution, we train an X-MACE model on molecular dipole moments by introducing geometry-dependent pseudo-partial charges, $q_i(\mathbf{R})$, assigned to each atom. Details on the architecture used can be found in the methods section. Briefly, the partial atomic charge is extracted as a latent property by learning molecular dipole moments via the following relation, where $q_i$ is the charge of atom a and $r_i$ is the distance vector of the atom from the molecule's centre of mass:\cite{gastegger2017machine, westermayr2022deep}
\begin{equation}
    \mu_i = \sum^{N_i}_i q_{i} r_i
\end{equation} 
For each chromophore variant, we compute a single scalar descriptor by averaging the predicted pseudo-charges over all atoms in the HBDI$^-$ backbone. The reddish radar plot in Fig.~\ref{fig:overview}h illustrates the results. As shown, increased conjugation allows the charge to delocalize over a larger portion of the molecule, reducing localized partial charges compared to HBDI$^-$ (with $R_5$ = CH$_3$). 
Tail-substituted systems cluster closer to zero, consistent with the negative charge on the oxygen being distributed over the extended $\pi$-system rather than remaining concentrated on the HBDI$^-$ backbone.

This change in charge distribution is directly reflected in the nature of the excited state. Photoexcitation to $S_1$ is dominated by a $\pi \rightarrow \pi^*$ transition (see orbitals in Fig.~\ref{fig:validation}c), which reduces the bond order along the methine bridge and softens the torsional coordinate, making twisting motions more accessible. However, increased conjugation counteracts efficient twisting in the excited state because rotation about either the imidazolinone or phenolate dihedral disrupts a larger amount of delocalization in more conjugated variants. As a result, highly conjugated chromophores incur a larger energetic penalty upon torsion, leading to an increased excited-state torsional barrier, as can be seen in the blue radar plot. This raised barrier slows access to the $S_1/S_0$ conical intersection at the torsional structure and suppresses non-radiative decay through photoisomerisation pathways. Consequently, population is retained on $S_1$ for longer times, increasing the likelihood of radiative relaxation and thereby favouring fluorescence. The opposite behaviour is observed for less conjugated HBDI$^-$ derivatives, reduced conjugation lowers the excited-state torsional barrier, enabling faster access to the conical intersection region. In this regime, torsion-driven internal conversion and isomerisation dominate the relaxation dynamics, leading to shorter excited-state lifetimes and higher photoisomerisation yields at the expense of fluorescence.

\begin{figure*}[htbp]
\centering
\includegraphics[width=\textwidth]{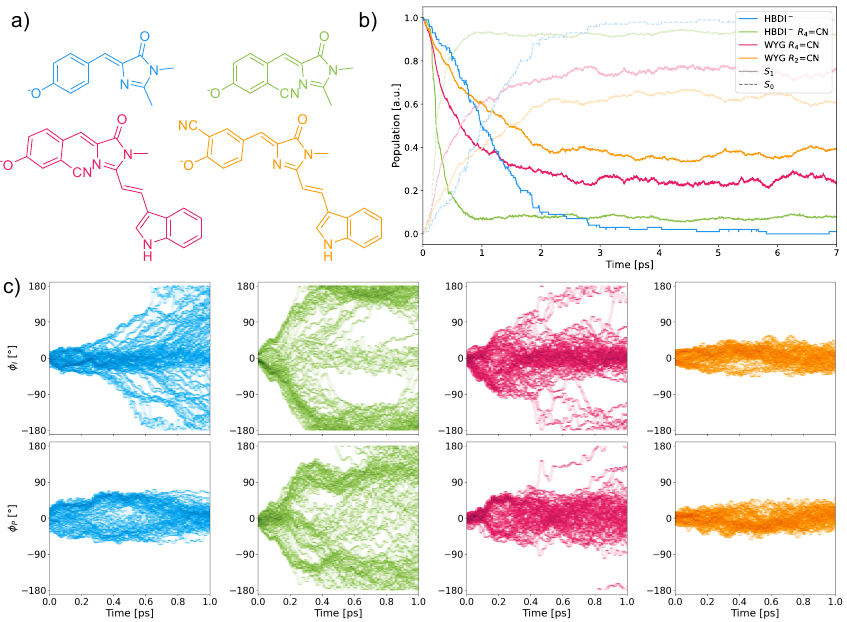}
\caption{Analysis of dynamical features of four exemplary HBDI$^-$ variants, consisting of pure HBDI$^-$ (blue), HBDI$^-$ with cyano-group substitution at $R_4$ (green), and HBDI$^-$ substituted with a WYG-like tail on $R_5$, as well as a cyano-group on $R_4$ (pink) or $R_2$(orange). panel a) shows the molecular structures. Panel b) depicts the population curves, with $S_0$ being shown in lighter colours than $S_1$ population. Panel c) shows the first picosecond of each trajectory as a function of dihedral angles $\phi_I$ (top) and $\phi_P$ (bottom),illustrating the torsion behaviour of the molecules.}
\label{fig:dynamics}
\end{figure*}
\subsection{Representative Lifetime Modifications and Trend Verification}
Taken together, the results observed by screening the excited-state dynamics of substituted HBDI$^-$ derivatives indicate two competing effects: (i) steric acceleration of torsion induced by substitutions on the phenyl ring, particularly at R$_4$, and (ii) electronic stabilization of the planar configuration arising from $\pi$-extension at the imidazole ring. To verify and illustrate these effects and their interplay, four prime examples are selected and compared, each representing a specific substitution pattern, illustrated in Figure \ref{fig:dynamics}a. As can be seen, we compare HBDI$^-$ to the following variants: HBDI$^-$ with a CN group at R$_4$ and  HBDI$^-$ with a WYG group at R$_5$ and a CN group at R$_2$ and R$_4$ individually. The corresponding population curves that show the amount of population in each state at a given time, extracted from the dynamics simulations, are visualized in panel b.
As can be seen, substitution of HBDI$^-$ only at the R$_4$ position with a CN group (green curves) leads to a steep initial decay of the S$_1$ population, reflecting efficient deactivation along the torsional coordinate. 
After this early decay, a fraction of trajectories that remain near-planar become trapped on the S$_1$ state, leading to a persistent excited-state population of ~10\% until the end of the 7~ps simulation. The population transfer from S$_1$ to S$_0$ is faster compared to HBDI$^-$ (blue curves).
This trend is visibly altered upon introducing additional substitution at the R$_5$ position of the imidazole ring (orange and pink curves). As can be seen, the population transfer is slower leading to a lifetime extension and an $S_1$ trapping of about 30\%. The resulting electronic effects become even more pronounced, when the phenolate substitution doesn't clash sterically, as can be seen for the $R_2$ substituted variant containing a WYG like tail (orange trajectories). Conjugated substituents at R$_5$ effectively enlarge the $\pi$-system of the chromophore, stabilizing the planar excited-state geometry and increasing the barrier along the torsional coordinates.

To verify these effects and highlight the impact of steric crowding, the trajectories dihedral angles $\phi_P$ and $\phi_I$ evolution are compared in panel c. As can be seen for the unsubstituted HBDI$^-$, the first $\sim$400~fs show only limited structural motion, consistent with the absence of a strong steric driving force. Only after several hundred femtoseconds does the imidazole ring begin to undergo the characteristic out-of-plane rotation typical for monomethine dyes. In contrast, the R$_4$ cyano-substituted variant (green lines), where the substituent sterically clashes with the imidazole ring, displays rapid and pronounced torsion; Both the imidazole and phenolate rings twist out of plane almost immediately, highlighting the sterically induced lowering of the torsional barrier and the resulting acceleration of the isomerisation and rapid de-excitation. 

For the WYG and R$_4$ CN-substituted HBDI$^-$ (pink lines), the initial torsional response remains visible. However, in many trajectories the dihedral angle approaches $\sim$90° before returning toward planarity rather than proceeding to full isomerisation, while many others stay planar. This behaviour, while mainly electronic in nature through the extended $\pi$ conjugation and its resulting planar stability, is supported by early nonradiative decay; the system undergoes internal conversion to the ground state near the twisted geometry and subsequently relaxes back to the initial Z-isomer, while planar structures remain trapped in $S_1$, resulting in $\sim$30\% $S_1$ population at the end of the simulation. 
In contrast, addition of CN at R$_2$ instead of R$_4$ (orange lines) leads to almost no distortion of the structures along the trajectories.  The resulting angle distribution shows almost planar structures during the first picosecond of simulation time. About 40\% of the trajectories remain in $S_1$ until the end of the simulation time.

Overall, our results identify a simple mechanistic principle governing the excited-state dynamics of substituted HBDI$^-$ derivatives. Steric crowding on the phenolate ring, most strongly at $R_4$, lowers the barrier to torsion and drives rapid access to twisted geometries that facilitate photoisomerisation. In contrast, $\pi$-extended substituents on the imidazole ring electronically stabilize the planar excited-state structure, oppose torsional progression, and promote recovery of the initial Z-isomer following internal conversion. The photochemical behaviour of each derivative is therefore governed by the competition between steric bias toward torsion and electronic bias toward planarity. These results show that increased $\pi$-conjugation is the primary driver of enhanced fluorescence, as it stabilizes the planar excited state, suppresses non-radiative decay pathways, and prolongs excited-state lifetimes.

\subsection{Conclusion}

In this work, a transferable and data-efficient route to excited-state dynamics screening is developed by combining the X-MACE excited-state model with curvature-driven surface hopping molecular dynamics simulations. Across a chemically diverse set of HBDI$^-$ derivatives, X-MACE remains reliable in the non-equilibrium, highly distorted regions that govern non-radiative decay, where transferability is typically most challenging. Remarkably, a single pretrained representation can be adapted using fewer than 100 reference geometries per variant while preserving key dynamical observables, including population decay, lifetimes and photoisomerisation yields. Together, these results elevate excited-state screening from static single-point analysis to statistically robust trajectory ensembles at scale.

Beyond enabling throughput, the screening reveals clear mechanistic structure property relationships that rationalize how substitution reshapes the photophysics of the GFP chromophore. Steric crowding on the phenolate ring, most prominently at R$_4$, lowers the effective torsional barrier, and accelerates access to twisted geometries, thereby shortening excited-state lifetimes and enhancing isomerisation yield. In contrast, $\pi$-extension at the imidazolinone side (R$_5$) stabilizes near-planar excited-state configurations by increasing the energetic penalty for torsion, suppressing efficient access to the conical intersection region and prolonging $S_1$ population. The resulting excited-state behaviour is governed by a competition between steric acceleration of twisting and electronic stabilization through conjugation, providing an interpretable design rule for tuning fluorescence versus non-radiative decay in green fluorescent protein (GFP) chromophore family.

More broadly, the strategy introduced here, pretraining on diverse excited-state data followed by coordinate-targeted, low-data fine-tuning, offers a general template for photochemical design. For HBDI$^-$, torsion and pyramidalisation provide the physically dominant coordinates for excited state processes; for other chromophore families, the same workflow can be coupled to system-specific coordinates or adaptive sampling when relaxation pathways are less well established. In this way, our results position transferable ML-driven dynamics not only as a practical accelerator, but as a route to mechanistic insights from large-scale excited-state trajectory data, enabling comparative screening of lifetimes and quantum yields at a scale that is effectively inaccessible to conventional electronic-structure dynamics. This capability extends structure-property understanding for broad, application critical chromophore classes such as GFP, where improved brightness and photostability directly enhance high-resolution imaging of living systems and the study of complex disease processes, including cancer progression and metastasis. Therefore, it turns excited-state dynamics into an actionable design engine, one that can move from observing photophysics to engineering it.


\section{Method}

\section{X-MACE Architecture}
Message Passing Neural Networks (MPNNs) \cite{gilmer2020message} are a family of neural architectures designed for graph-structured data. They are widely used in molecular modeling because they can encode both local and long-range structural information through iterative updates of node representations \cite{gilmer2017neural, yang2019analyzing}. In molecular graphs, nodes usually correspond to atoms, while edges represent chemical bonds or other interactions. A molecular system is represented as a graph \( G = (V, E) \), where \( V \) denotes the set of nodes and \( E \) the set of edges. For each node \( i \in V \), the associated feature tuple at layer \( t \) is

\begin{align}
    \sigma_i^{(t)} = \left( \mathbf{h}_i^{(t)}, Z_i, \mathbf{R}_i \right),
\end{align}

where \( \mathbf{h}_i^{(t)} \) is the hidden feature vector of node \( i \) at iteration \( t \), \( Z_i \) is the atomic number, and \( \mathbf{R}_i \) is the three-dimensional position vector of atom \( i \). An MPNN consists of two main stages: an MP stage and a readout stage. In the MP stage, node features are updated for \( n \) iterations such that the graph structure is progressively encoded into the node states. At iteration \( t \), each node \( i \) receives an aggregated message \( \mathbf{m}_i^{(t)} \) from its neighbors \( \mathcal{N}(i) \):

\begin{align}
    \mathbf{m}_i^{(t)} = \sum_{j \in \mathcal{N}(i)} M\left( \mathbf{h}_i^{(t)}, \mathbf{h}_j^{(t)}, \mathbf{r}_{ij} \right),
\end{align}

where \( M \) is the message function, and \( \mathbf{r}_{ij} = \mathbf{R}_j - \mathbf{R}_i \) is the relative position vector between atoms \( i \) and \( j \). The hidden state of node \( i \) is then updated through an iteration-dependent update function \( U_t \):

\begin{align}
    \mathbf{h}_i^{(t+1)} = U_t\left( \mathbf{h}_i^{(t)}, \mathbf{m}_i^{(t)} \right).
\end{align}

By repeating this process, information propagates through the graph, allowing each node to build a richer representation of its local environment. To model interactions beyond pairwise terms, an atomic cluster expansion (ACE) \cite{dusson2022atomic, ho2024atomic} can be incorporated into the MPNN framework. This enables the inclusion of many-body contributions, which are important for accurately describing complex molecular interactions. Combining ACE with MP yields the MACE model \cite{batatia2022mace}. With this extension, the message \( \mathbf{m}_i^{(t)} \) at iteration \( t \) can be written as a sum of many-body interaction terms:

\begin{align*}
    \mathbf{m}_i^{(t)} &= \sum_{j \in \mathcal{N}_1(i)} u_1 \left( \sigma_i^{(t)}, \sigma_j^{(t)} \right) \\
    &+ \sum_{j_1 \in \mathcal{N}_2(i)} \sum_{j_2 \in \mathcal{N}_2(j_1)} u_2 \left( \sigma_i^{(t)}, \sigma_{j_1}^{(t)}, \sigma_{j_2}^{(t)} \right) \\
    &\quad \vdots \\
    &+ \sum_{j_1, \dots, j_\nu \in \mathcal{N}_\nu(i)} u_\nu \left( \sigma_i^{(t)}, \sigma_{j_1}^{(t)}, \dots, \sigma_{j_\nu}^{(t)} \right),
\end{align*}

where \( u_\nu \) denotes the interaction function for the \( \nu \)-body term, and \( \mathcal{N}_\nu(i) \) represents the set of \( \nu \)-th order neighbor collections associated with node \( i \). A key requirement in physical modeling is that predictions transform correctly under rotations and translations. Accordingly, the messages are constructed to be equivariant with respect to the rotation group \( SO(3) \). In practice, this ensures that when atomic coordinates are rotated, predicted quantities such as forces and energies transform consistently. In the readout stage, the network aggregates information from the updated node states to produce a global molecular property. Each node contributes a scalar value, and the total prediction is obtained by summation:

\begin{align}
    \rho = \sum_{i \in V} R\left( \mathbf{h}_i^{(t)} \right),
\end{align}

where \( R \) is the readout function mapping the hidden state of node \( i \) to a scalar contribution, and \( \rho \) denotes a molecular property such as the total energy.

In this work, we extend MACE to X-MACE to predict multiple potential energy surfaces, corresponding forces and spectral properties. Therefore we have multiple readouts.

\subsection{X-MACE Charge Model}
To predict partial atomic charges that are not directly accessible from quantum-chemical calculations, we further adapted the X-MACE model. Therefore, in the readout stage, the network aggregates information from the updated node states to produce molecular-level predictions. For scalar extensive properties, each node contributes a scalar value, and the total prediction is obtained by summation:

\begin{align}
    \rho = \sum_{i \in V} R\left( \mathbf{h}_i^{(t)} \right),
\end{align}

where \( R \) is the readout function mapping the hidden state of node \( i \) to a scalar contribution, and \( \rho \) denotes a molecular property such as the total energy.

In our case, the readout is instead used to predict atom-centred pseudo-charges \(q_i(\mathbf{R})\), which are then combined to reconstruct the molecular dipole. Specifically, the network assigns a scalar pseudo-charge to each atom through an atom-wise readout, and the molecular dipole is obtained as the charge-weighted sum over atomic positions:

\begin{equation}
\boldsymbol{\mu}(\mathbf{R})=\sum_{i=1}^{N} q_i(\mathbf{R})\,\mathbf{r}_i ,
\label{eq:dipole_pseudocharges}
\end{equation}

where \(\mathbf{r}_i\) denotes the position of atom \(i\). This readout preserves the atom-wise decomposition while yielding a physically meaningful vector observable. For each chromophore variant, we further define a single scalar descriptor by averaging the predicted pseudo-charges over all atoms in the HBDI$^-$ backbone:
\begin{equation}
\bar{q}_{\mathrm{HBDI}^-}=\frac{1}{N_{\mathrm{bb}}}\sum_{i \in \mathcal{B}} q_i(\mathbf{R}),
\end{equation}
where \(\mathcal{B}\) denotes the set of backbone atoms and \(N_{\mathrm{bb}} = |\mathcal{B}|\).

\subsection{Quantum Chemical Reference Calculations}
To identify a suitable electronic structure method for non-modified HBDI$^-$, we compared second order perturbation theory (ADC(2))\cite{dreuw2015algebraic} with complete active space self consistent field (CASSCF)\cite{roos1980casscf} calculations employing different active spaces (4,3), (10,8), and (16,14). HBDI$^-$ was chosen as a benchmark system because high-level \textit{ab initio} reference data and experimental results are available for validation. Single-point calculations were performed for nine representative geometries along the relevant torsional pathways and compared to SA3-XMS-CASPT2 reference energies obtained from Ref.\cite{list2022internal}. All tested methods reproduced the ground-state (S$_0$) potential energy surface satisfactorily. However, only ADC(2) accurately reproduced the shape of the S$_1$ potential energy surface along the torsional coordinates (see Figure~\ref{fig:validation}a).

Inspection of the CASSCF active spaces and their orbitals, which were generated in Orca,\cite{neese2025software} revealed that key bonding, non-bonding, and antibonding orbitals localized on the methine bridge were not consistently included (See Section S2 im the SI). Reliable description of the excited-state surface would therefore require manual reordering and selection of active orbitals. While feasible for a single system, this procedure becomes impractical for nonadiabatic dynamics simulations and systematic screening of multiple derivatives.

ADC(2)/cc-pVDZ shows good agreement with SA3-XMS-CASPT2 reference data for the relevant excited-state energetics. To further validate the method, absorption spectra of HBDI$^-$ were computed by sampling 100 geometries from harmonic Wigner distributions at 0~K and 300~K, and compared to experimental spectrum from Nielsen \textit{et al.}\cite{nielsen2001absorption}(see Fig.~\ref{fig:validation}c).

The spectrum obtained from 0~K sampling exhibits a maximum at 2.7~eV and was shifted by $-0.1$~eV to align with the experimental absorption maximum. The overall spectral shape agrees well with experiment, particularly the low-energy features between 2.0 and 2.4~eV, although their intensities are slightly overestimated. The shoulder near 2.7~eV is not well reproduced at this temperature. When the sampling temperature is increased to 300~K to better mimic experimental conditions, the agreement improves. The absorption maximum requires only a small shift of $-0.02$~eV, and the slightly blue-shifted shoulder is reproduced more accurately. However, the low-energy fine structure becomes less pronounced, and an additional shoulder appears around 2.45~eV. These results indicate that the sampling temperature significantly influences the spectral lineshape, suggesting that thermally activated vibrational motion enables additional absorption pathways.

Overall, ADC(2)/aug-cc-pVDZ reliably reproduces both high-level \textit{ab initio} reference data and the experimental gas-phase absorption spectrum of HBDI$^-$. In contrast to CASSCF, ADC(2) does not require system-specific active space selection and can therefore be applied in a robust, black-box manner. For these reasons, ADC(2)/cc-pVDZ was chosen for the subsequent nonadiabatic dynamics simulations and screening calculations.

\subsection{Pre-Training Set Generation}
The chromophores dataset for transferable excited-state machine learning was generated with this study. A total of 368 SMILES strings were taken from Ref.\cite{joung2020experimental}. Initial 3D geometries were constructed using RDKit\cite{landrum2013rdkit} and then expanded as follows: Metadynamics simulations were performed using the xtb program, \cite{bannwarth2019gfn2} with a bias potential strength of 0.1405 Hartree (kpush) and a Gaussian width of 0.01125 (alp). The underlying molecular dynamics simulations were conducted at 300 K with a timestep of 1.0 fs, a total simulation time of 10 ps, and trajectory data recorded every 200 fs. The recorded data points were recomputed with the ADC(2) for the first 5 singlet states. A total of 12,183 calculations converged, resulting in the final data set.

\subsection{Fine-Tuning Data Generation}
To generate the fine-tuning dataset, we began with the HBDI$^-$ molecule as the parent scaffold. Substituted variants were created by modifying predefined positions on the phenolate and imidazolinone rings, denoted as $R_1$–$R_4$ and $R_5$, respectively. For each variant, a single substitution site was selected, and the atom at that position was replaced with a functional group, while the remainder of the HBDI$^-$ backbone was left unchanged.

The local attachment direction of each substituent was defined based on the scaffold geometry, using the vector between the backbone attachment atom and the atom being replaced. New functional groups were then positioned along this axis and rotated to achieve the desired bonding geometry. For substituents with multiple plausible torsional configurations (e.g., multi-atom carbonyl-containing groups), several candidate orientations were generated by rotation around the attachment axis. The configuration that maximized the minimum distance to existing HBDI$^-$ atoms was selected to minimize steric clashes.

Each substituted structure was subsequently relaxed via local geometry optimization using the GFN2-xTB forcefield\cite{bannwarth2019gfn2}, allowing the newly introduced atoms to adjust and resolving any unfavorable contacts arising from the initial placement. Each chromophore thus contained one substitution on the phenyl ring and, in some cases, an additional substitution at $R_5$.

Following this procedure, a fine-tuning dataset was created by sampling 90 datapoints per variant. These datapoints were selected along the two torsion pathways, recreating the torsional angles described in \cite{list2022internal}. To do so, the initially planar geometries of the 193 selected variants were twisted to show the same dihedral angles as the nine reference geometries reported in \cite{list2022internal} using rdkit. Subsequently, the bridge hydrogen atoms were relaxed using the GFN2-xTB forcefield\cite{bannwarth2019gfn2} for 1000 steps, while the remaining atoms were fixed, to avoid replanarisation.  The obtained molecules were used to perform Wigner sampling at 300 K, creating 1000 conformations, of which 10 were randomly sampled and labelled using ADC(2)/aug-cc-pVDZ. The labelled structures were aggregated to form the training data set. 

\subsection{Model Performance}
\begin{figure*}[htbp]
    \centering
    \includegraphics[width=\textwidth]{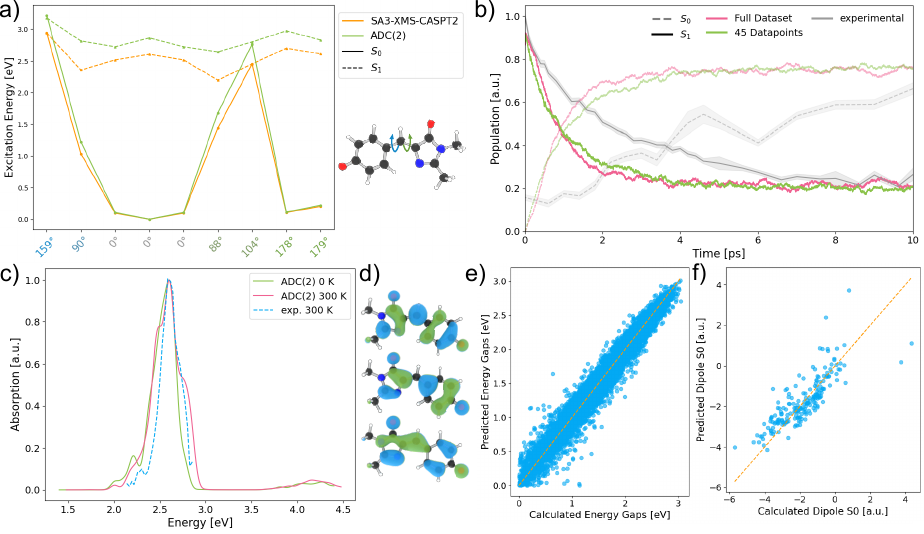}
    \caption{(a) Potential energy profiles along torsional coordinates comparing ADC(2) with SA3-XMS-CASPT2 reference method from List \textit{et al.}\cite{list2022internal}(b) comparison of excited-state population dynamics obtained from surface hopping simulations using the full reference dataset ($\sim2000$ datapoints) and a reduced fine-tuning set (45 datapoints), alongside experimental data from Svendsen \textit{et al.}\cite{svendsen2017origin}. (c) Absorption spectra computed at the ADC(2) level at 0 K and 300 K compared to experimental spectrum\cite{nielsen2001absorption}. (d) Representative molecular orbitals illustrating the bonding, non-bonding and anti-bonding orbitals from bottom to top ($\pi$ → $\pi*$ excitation from middle to top). (e) Scatter plot of predicted versus reference energy gaps. (f) Correlation between predicted and reference dipole moments.}
    \label{fig:validation}  
\end{figure*}

Scatter plots comparing reference and predicted energy gaps across all investigated chromophore variants (Fig. \ref{fig:validation}e) demonstrate good agreement over the full energy range. Importantly, this analysis includes not only unsubstituted HBDI$^-$ but also all substituted derivatives, confirming that the model generalizes across chemical modifications. The accurate reconstruction of energies ensures that the machine-learned potential preserves the distances between the underlying potential energy surfaces and therefore provides a robust foundation for nonadiabatic molecular dynamics simulations.

Nonadiabatic molecular dynamics simulations with population curves shown in panel b of Figure \ref{fig:validation} that are based on the refined model (green) reproduce the experimentally observed population decay timescales from pump–probe measurements for HBDI$^-$ (grey)\cite{svendsen2017origin}, providing additional validation at the dynamical level. Notably, fine-tuning the pre-trained model with only 45 HBDI$^-$ data points achieves comparable accuracy to training a model from scratch on approximately 2000 HBDI$^-$ reference configurations, demonstrating the transferability and data efficiency of the approach.

Together, these results establish that the chosen electronic structure method and machine-learned potential provide a reliable description of both the static and dynamical photophysics of HBDI$^-$.

\subsection{Oscillator Strength Prediction}
To enable direct prediction of absorption properties we trained X-MACE to additionally predict electronic transition oscillator strengths alongside state energies and forces. For each geometry $\mathbf{R}$, the model outputs oscillator strengths $f_{0n}(\mathbf{R})$ for transitions from the ground state $S_0$ to excited states $S_n$ included in the training set. These quantities are learned as scalar targets using an atom-wise readout followed by pooling to a molecular prediction, analogous to other global properties in the MACE framework.

In the reference calculations, oscillator strengths were obtained consistently with ADC(2) used to generate the excited-state energies and gradients. During training, oscillator strengths were included as an additional supervised target. Following fine-tuning, predicted oscillator strengths were used (i) to construct absorption spectra for each variant, and (ii) to define excitation conditions for the subsequent nonadiabatic molecular dynamics simulations by selecting initial electronic states according to the relative transition probabilities.

\subsection{Surface Hopping Molecular Dynamics}
The surface hopping dynamics were performed using the SHARC dynamics suite,\cite{mai2018nonadiabatic} that was interfaced with X-MACE in this work. The dynamics were propagated for 7~ps using 0.5~fs time steps and 25 substeps each. During the propagation, the diagonal representation was used and nonadiabatic couplings were approximated using curvature driven time-derivative couplings (kTDC)\cite{shu2022nonadiabatic}. Further information on the SHARC dynamics setup, as well as the creation of the simulations initial conditions can be found in the section S5 of the SI. 
The creation of initial conditions usually requires costly quantum-chemical calculations of excitation energies and oscillator strengths for a variety of geometries. This step is further accelerated using X-MACE to predict the needed properties. 

\section{Acknowledgements}
Financial support by the Deutsche Forschungsgemeinschaft (DFG, German Research Foundation) – TRR 325 (project C7) – 444632635 and RTG 2721 is gratefully acknowledged. Computations for this work were done using resources of the Leipzig University Computing Center, the NHR Center at TU Dresden, and Paderborn Center for Parallel Computing (PC2). S.W. and J. W. are supported by a pre-doc award from Leipzig University. 

\section*{Competing Interests}
The authors declare no competing interests.

\section{Code Availability}
The associated X-MACE code can be found at https://github.com/rhyan10/X-MACE

\section{Data Availability}
All data and additional helper scripts for this work can be found at \url{https://figshare.com/articles/dataset/Data_and_code_zip_folder_for_X-MACE/31871941}

\section{extended data figure}

\FloatBarrier

\section*{Supplementary Information}
\section{Structure Construction Details}
\begin{figure}[h]
    \centering
    \includegraphics[width=0.5\linewidth]{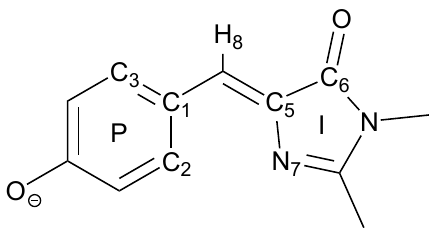}
    \caption{Molecular structure of HBDI$^-$ with atom position definitions used in Eq.~\ref{molecules-start}-\ref{molecules-end}}
    \label{fig:atoms-def}
\end{figure}

The GFP chromophore HBDI$^-$ (see Figure~\ref{fig:atoms-def}) consists of two conjugated rings connected by a methine bridge, forming a $\pi$-system that stretches over nearly the whole molecule.
This high delocalization of charges makes the molecule very rigid in the ground state. 

Upon excitation, the rotation of either of the two rings is enabled, through which a CI can be reached.
This makes the change of the dihedral angles a key factor for excited-state dynamics of HBDI$^-$.
Coupled to these dihedral angle changes is the pyramidalization of the methine bridge.
For this reason the dihedral ($\phi_I$, $\phi_P$) and pyramidalization ($\theta_{pyr}$) angles are tracked throughout this work according to formulas \ref{molecules-start} to \ref{molecules-end}. \cite{list2022internal}

\begin{equation}\label{molecules-start}
    \theta_{pyr} = arccos((e_{C_1-C_4})\cdot e_{H_8-C_4})-\frac{\pi}{2}
\end{equation}

\begin{equation}
    \phi_I = sgn_I\cdot arccos((e_{C_1-C_4} \times e_{C_5-C_4}) \cdot (e_{C_5-C_4} \times e_{C_6-N_7}))
\end{equation}
with:
\begin{equation}
    sgn_I = sgn(((e_{C_1-C_4}\times e_{C_5-C_4}) \times (e_{C_5-C_4} \times e_{C_6-N_7}))\cdot e_{C_5-C_4})
\end{equation}

\begin{equation}
    \phi_P = sgn_P\cdot arccos((e_{C_5-C_4} \times e_{C_1-C_4}) \cdot (e_{C_1-C_4} \times e_{C_3-C_2}))
\end{equation}
with:
\begin{equation}\label{molecules-end}
    sgn_P = sgn(((e_{C_5-C_4})\times(e_{C_1-C_4}) \times (e_{C_1-C_4} \times e_{C_3-C_2}))\cdot e_{C_1-C_4})
\end{equation}

Throughout the work geometries along the torsional coordinates are used. These geometries were adapted from List \textit{et al}\cite{list2022internal}, to compare different levels of theory to their reported SA3-XMS-CASPT2(4,3)/6-31G* calculations. the structures were further used to create training datapoints, by adapting the dihedral angles reported in table~\ref{tab:dihedral_angles}.
\begin{table}[h]
\centering
\caption{Dihedral angles $\phi_I$ and $\phi_P$ (in degrees) for the investigated stationary points and conical intersections.}
\vspace{2mm}
\setlength{\tabcolsep}{4pt}
\small
\begin{tabular}{lrrrrrrrrr}
\hline
 & MECI-I & MECI-I-2 & S$_0$ min & S$_0$ min-E & S$_1$-I & S$_1$-P & S$_1$-planar & S$_1$-planar-E & MECI-P \\
\hline
$\phi_I$ (°) & 104.63 & 71.38 & 0.03 & 179.89 & 88.64 & -0.04 & 0.07 & 178.85 & -53.70 \\
$\phi_P$ (°) & -30.55 & 29.94 & 0.06 & 0.01 & -1.62 & -90.23 & -0.11 & -0.42 & 96.71 \\
\hline
\end{tabular}
\label{tab:dihedral_angles}
\end{table}

\section{Quantum Chemical Method Comparison}

\begin{figure}
    \centering
    \includegraphics[width=0.95\linewidth]{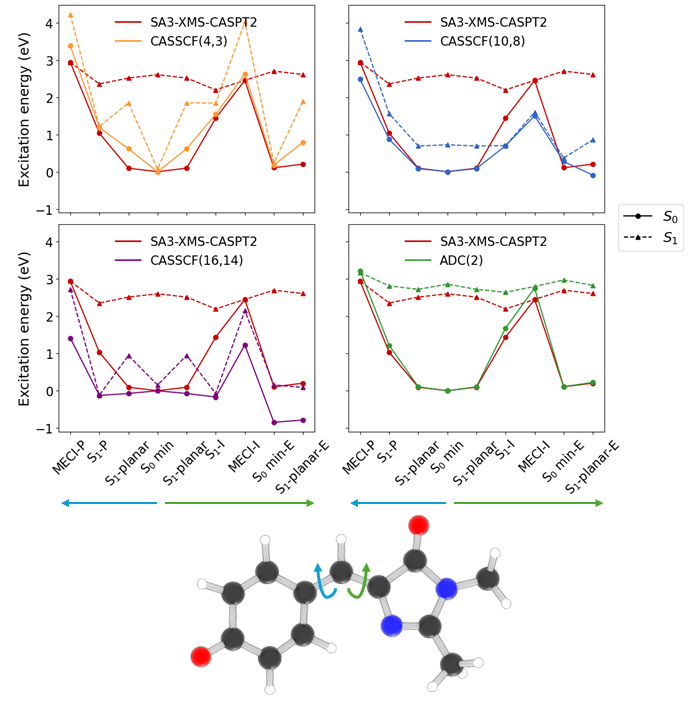}
    \caption{Ground state energies (S$_0$, solid lines) and vertical excitation energies (S$_1$, dashed lines) for eight different geometries of HBDI$^-$, with increasing torsion angles from center to outside. Energies for CASSCF calculations with active spaces (3,4), (10,8), and (16,14) and ADC(2) are shown additionally to values calculated using CASPT2 from \cite{list2022internal}}
    \label{fig:vertical-excitation-energies}
\end{figure}

\begin{figure}
    \centering
    \includegraphics[width=1\linewidth]{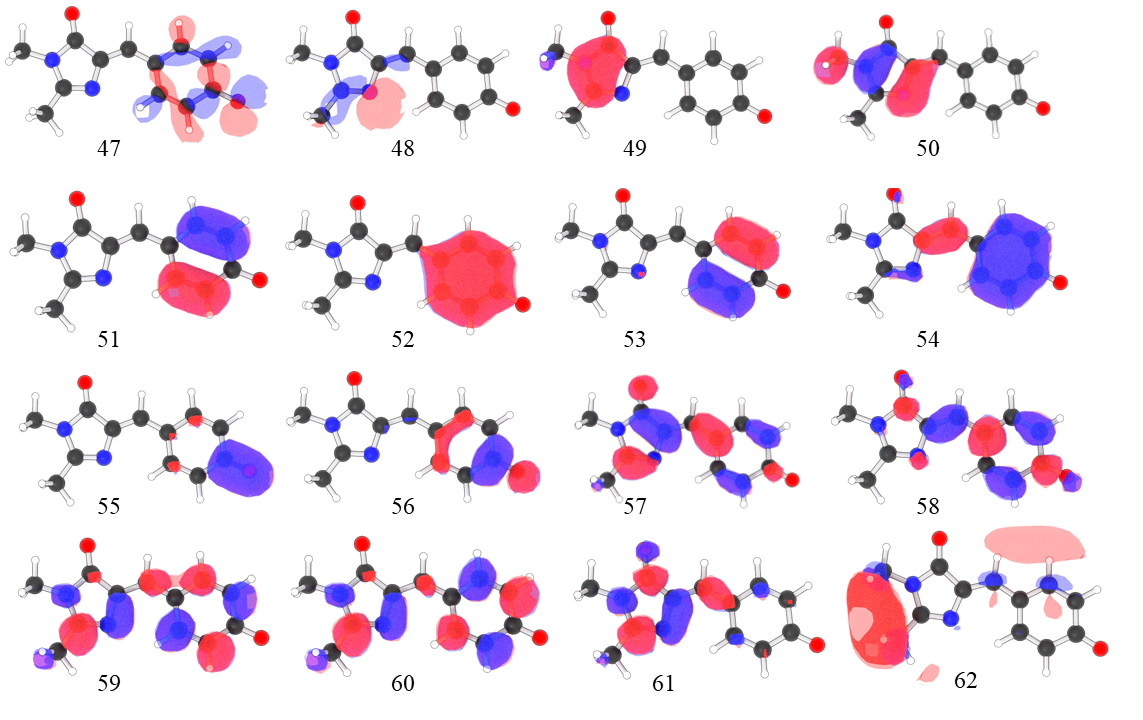}
    \caption{Orbitals of HBDI$^-$ considered for the CASSCF calculations. For the CASSCF(4,3) calculation, orbitals 54-56 were located in the active space. For CASSCF(10,8) orbitals 51-58, and for CASSCF(16,14) orbitals 48-61 were located in the active space.}
    \label{fig:CASSCF-orbitals-all}
\end{figure}

As explained in the main text, the quantum chemical method was chosen by comparing different levels of theory.
To do so, different CASSCF active spaces were compared to CASPT2 single point energies reported by List \textit{et al.}\cite{list2022internal} along the torsional pathways (Fig.\ref{fig:vertical-excitation-energies}).
To explain the poor fit between the CASSCF and CASPT2 results from List \textit{et al.}\cite{list2022internal}, the orbitals inside the active spaces are shown in Fig~\ref{fig:CASSCF-orbitals-all}. It can be seen, that the relevant non- and antibonding orbitals of the methine bridge are not included in the active spaces,making CASSCF(4,3)/cc-pVDZ. CASSCF(10,8)/cc-pVDZ and CASSCF(16,14)/cc-pVDZ not suitable for the reported study.

\section{Model Training}
The model was initially trained on the chromophore dataset and initialized with the MACE-OFF\cite{kovacs2025mace} parameters. In this case the output layers were randomly initialized. We then trained an X-MACE model by fine-tuning this pretrained chromophore model on our HBDI$^-$ variants dataset. To ensure reproducibility, we fixed the random seed to 100 and reserved 10\% of the data for validation. The model was configured to predict three excited-state energies and forces, although only two of these were used in the present work. We used a local atomic neighbourhood cutoff of 5.0~\AA, two message-passing interaction layers, and a third-order correlation setting to capture many-body interactions. The hidden representation was defined by \(128\mathrm{x}0e + 128\mathrm{x}1o\), with an MLP readout of \(128\mathrm{x}0e\) and 8 radial basis functions. Training was performed for up to 100 epochs with a batch size of 100, a learning rate of 0.001, single-precision arithmetic, and GPU acceleration. An exponential moving average (EMA) of the model weights was enabled with a decay factor of 0.99 to improve training stability. Both the energy and force terms in the loss were the mean squared error and weighted by 100. The same training hyperparameters were used for each model.

The charge model was trained on the HBDI$^-$ dipoles calculated at the ADC(2)/aug-cc-pVDZ level of theory. 10\% of the data was removed to form a validation set. The model was trained to predict dipoles for the first 3 states, using a local atomic neighborhood cutoff of 5.O~\AA, two message passing interaction layers and a third-order correlation setting. The training was performed for 100 epochs, using a learning rate of 0.0001 and an exponential moving average with a factor of 0.99. To evaluate the trained model, calculated dipoles were plotted against predicted values (extended data figure of the manuscript), showing good agreement.

\section{Validation of chromophores and HBDI$^-$ fine-tuned models}

To validate the models for chromophore predictions and the fine-tuned HBDI$^-$ system, we compared reference forces and energies against those predicted by the machine-learning potential. For the chromophores dataset, reference forces for the ground state and first excited state, computed at the ADC(2)/TZVP level, as well as energy gaps between all electronic states relative to the ground state, were plotted against model predictions (Fig.\ref{fig:chromophores-scatterplots-energies} and Fig.\ref{fig:chromophores-scatterplots-forces}). In both cases, the predictions show good agreement with the reference data, although a noticeable spread is observed in the force predictions, indicating slightly larger deviations for this quantity.

\begin{figure}
    \centering
    \includegraphics[width=0.8\linewidth]{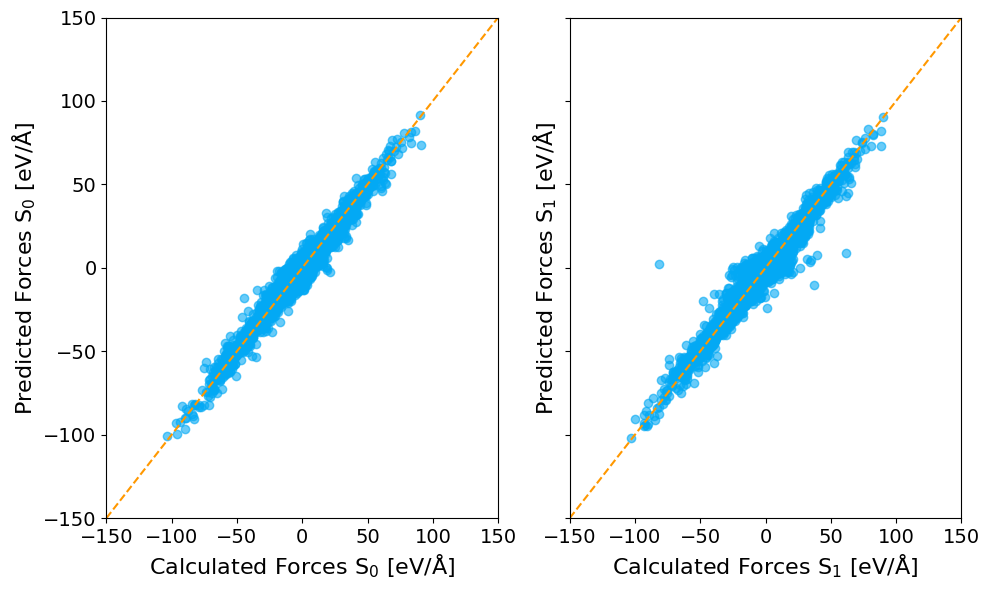}
    \caption{Parity plots of predicted versus calculated forces (in eV/\AA) for the HBDI system. The left panel corresponds to the ground state (S$_0$), while the right panel shows the first excited state (S$_1$). The dashed orange lines indicate perfect agreement ($y=x$).}
    \label{fig:scatterplots-forces}
\end{figure}

For the fine-tuned HBDI$^-$ model, performance was evaluated in more detail through comparisons of reference and predicted forces and energy gaps. Scatter plots comparing forces computed at the ADC(2)/aug-cc-pVDZ level with those predicted by the X-MACE model (Fig.\ref{fig:scatterplots-forces}) demonstrate strong correlation on the training data, confirming that the model reproduces reference forces with sufficient accuracy for stable dynamics. Prior to generating these plots, structures exhibiting unphysically large forces arising from atomic overlaps or extremely short interatomic distances in torsion. These geometries were filtered out. These configurations originate from the dataset construction, where torsional structures were intentionally not relaxed to prevent replanarization, occasionally leading to interatomic distances below 1 \AA. A cutoff of 1 \AA was therefore applied to exclude such clearly unphysical geometries. This filtering does not affect model training or its applicability, as these high-energy configurations are unlikely to be sampled during simulations.

\begin{figure}
    \centering
    \includegraphics[width=0.5\linewidth]{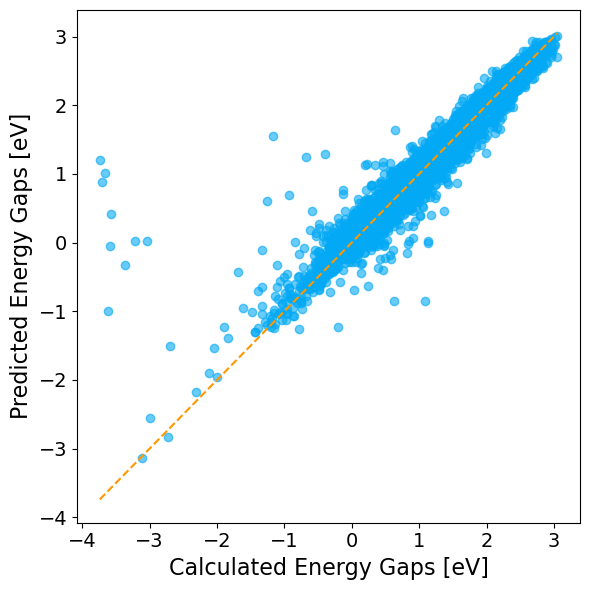}
    \caption{Parity plot of predicted versus calculated S$_1$–S$_0$ energy gaps (in eV). The dashed orange line represents perfect agreement ($y=x$).}
    \label{fig:scatterplots-energy}
\end{figure}

In addition to forces, model performance was assessed by comparing reference and predicted electronic energy gaps ($\Delta E = E_{S_1} - E_{S_0}$), shown in Fig.\ref{fig:scatterplots-energy}. These gaps are key quantities governing nonadiabatic dynamics, and the corresponding scatter plots reveal strong agreement between reference values and X-MACE predictions. While force-based validation ensures stable molecular dynamics, the accurate reproduction of energy gaps confirms that the model reliably captures the energetic landscape relevant for surface hopping simulations. 


\subsection{Breakdown of errors per molecule}
\begin{figure*}[p]
    \centering
    \includegraphics[page=1,width=\textwidth]{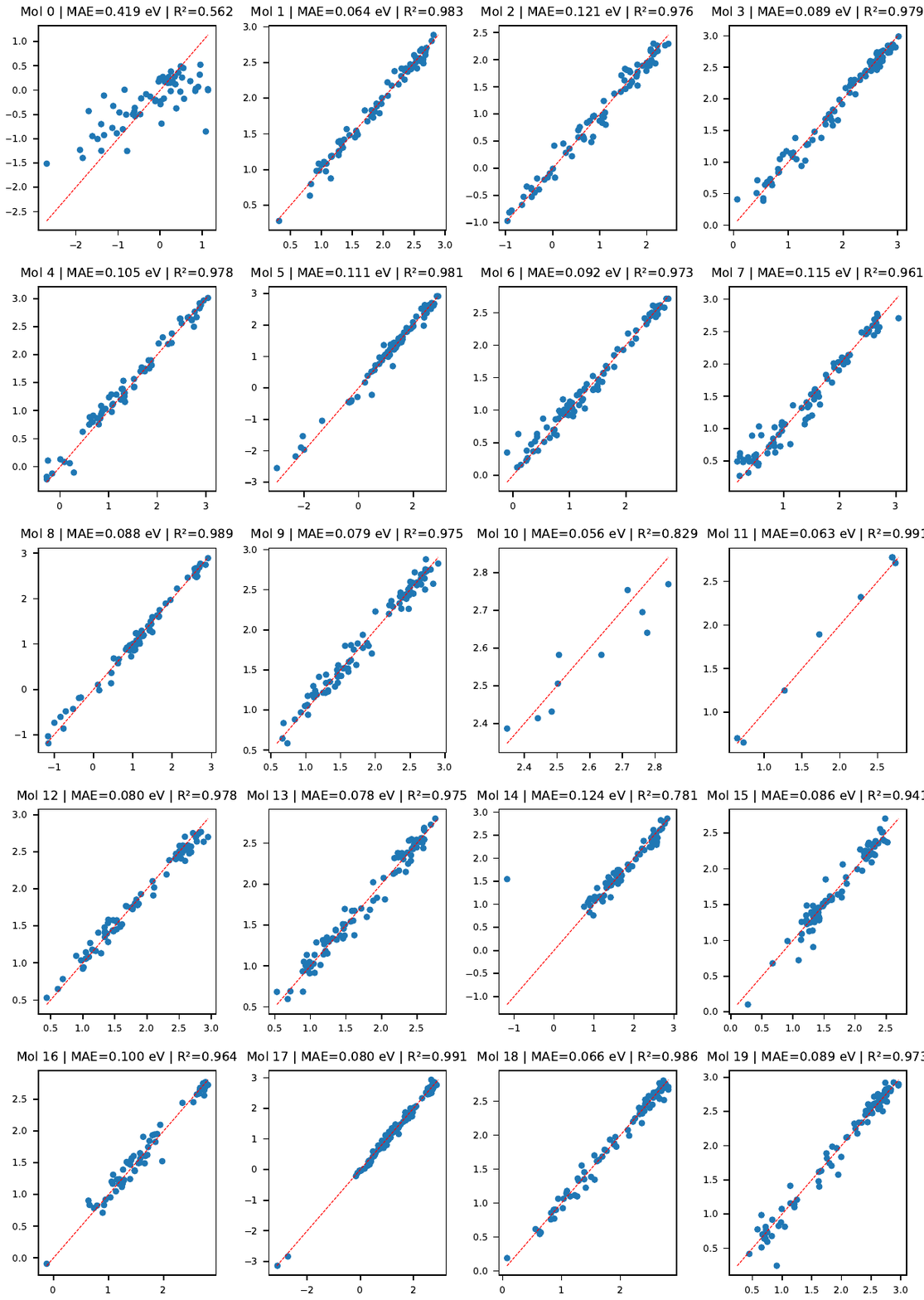}
\end{figure*}

\begin{figure*}[p]
    \centering
    \includegraphics[page=2,width=\textwidth]{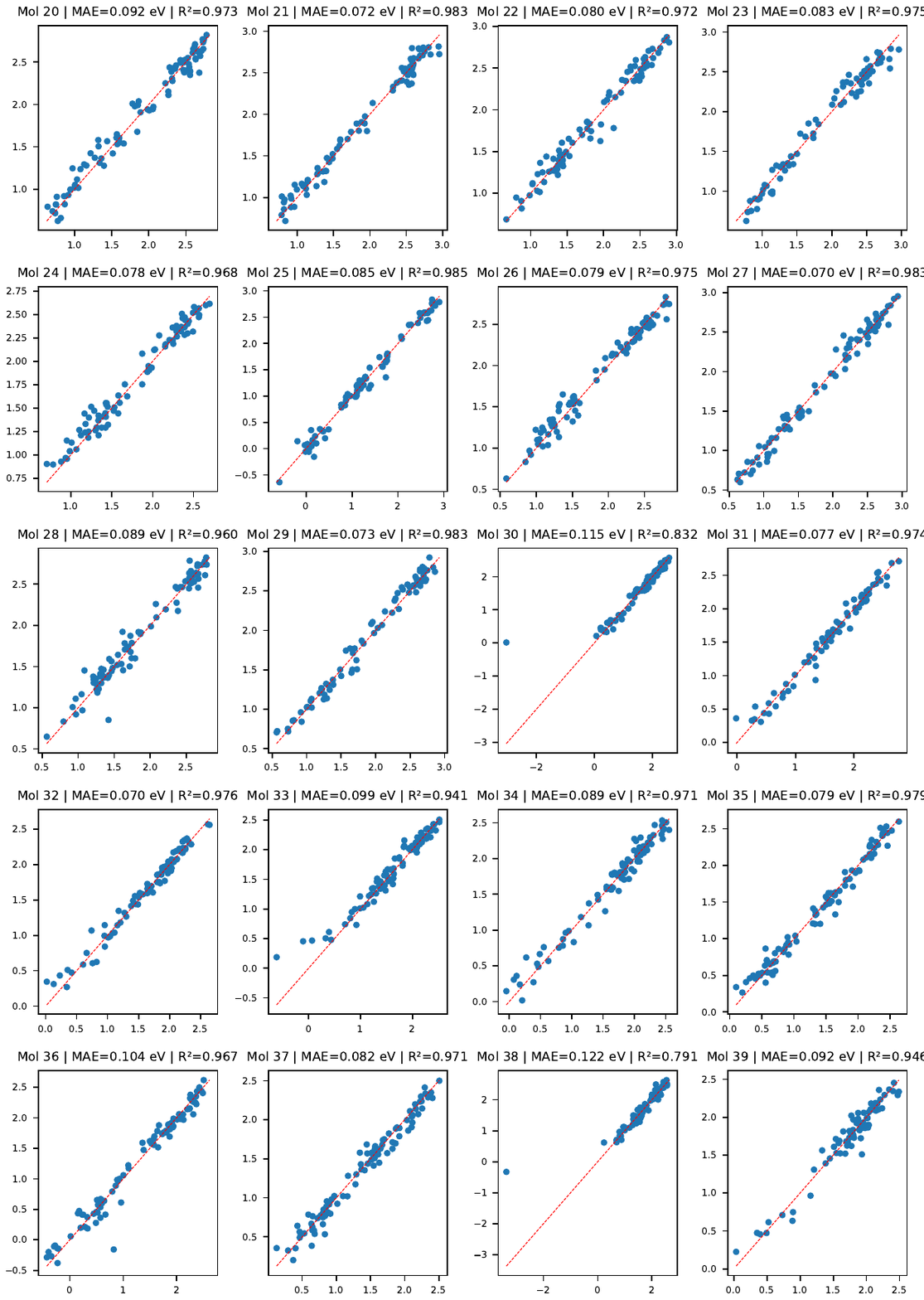}
\end{figure*}

\begin{figure*}[p]
    \centering
    \includegraphics[page=3,width=\textwidth]{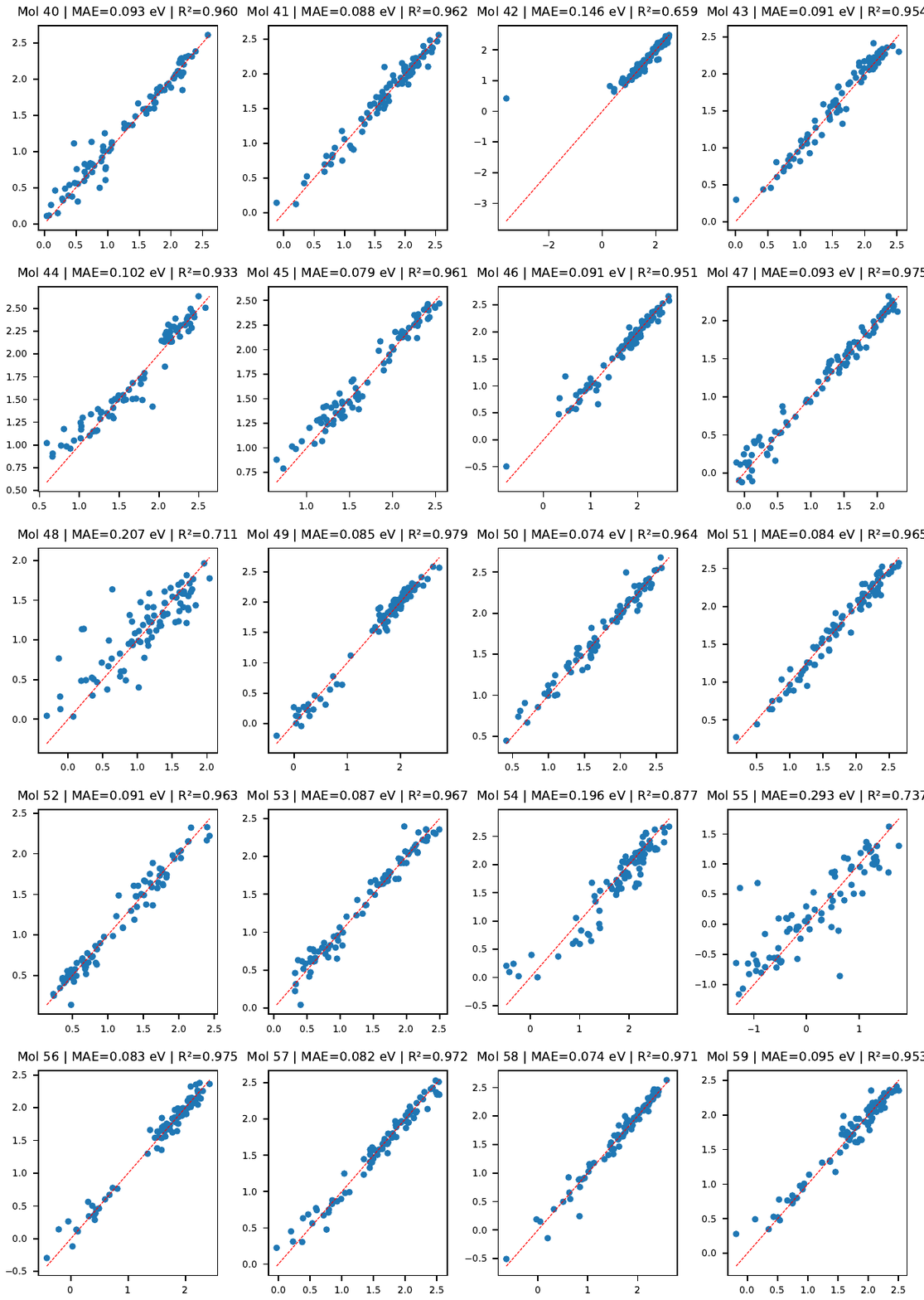}
\end{figure*}

\begin{figure*}[p]
    \centering
    \includegraphics[page=4,width=\textwidth]{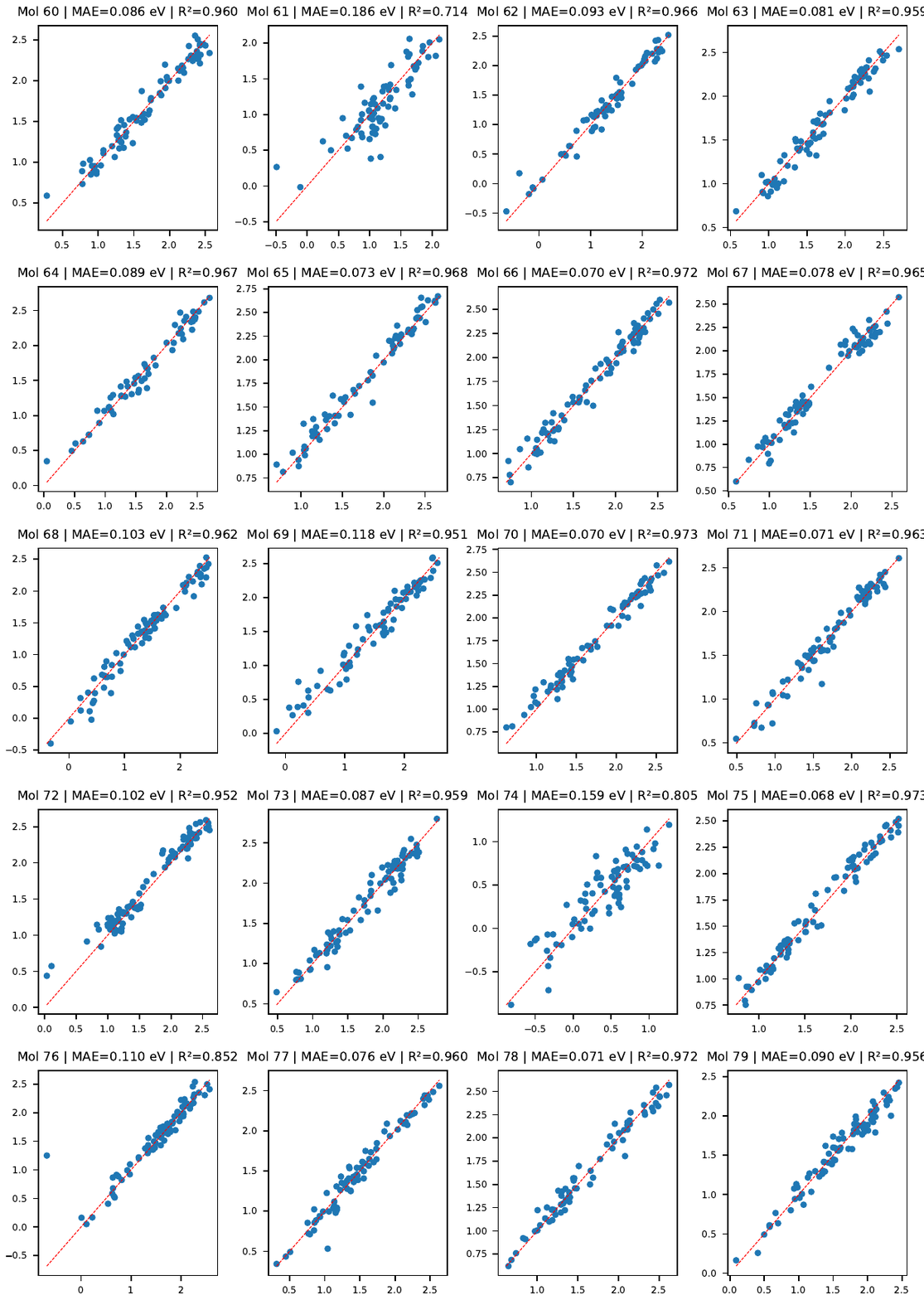}
\end{figure*}

\begin{figure*}[p]
    \centering
    \includegraphics[page=5,width=\textwidth]{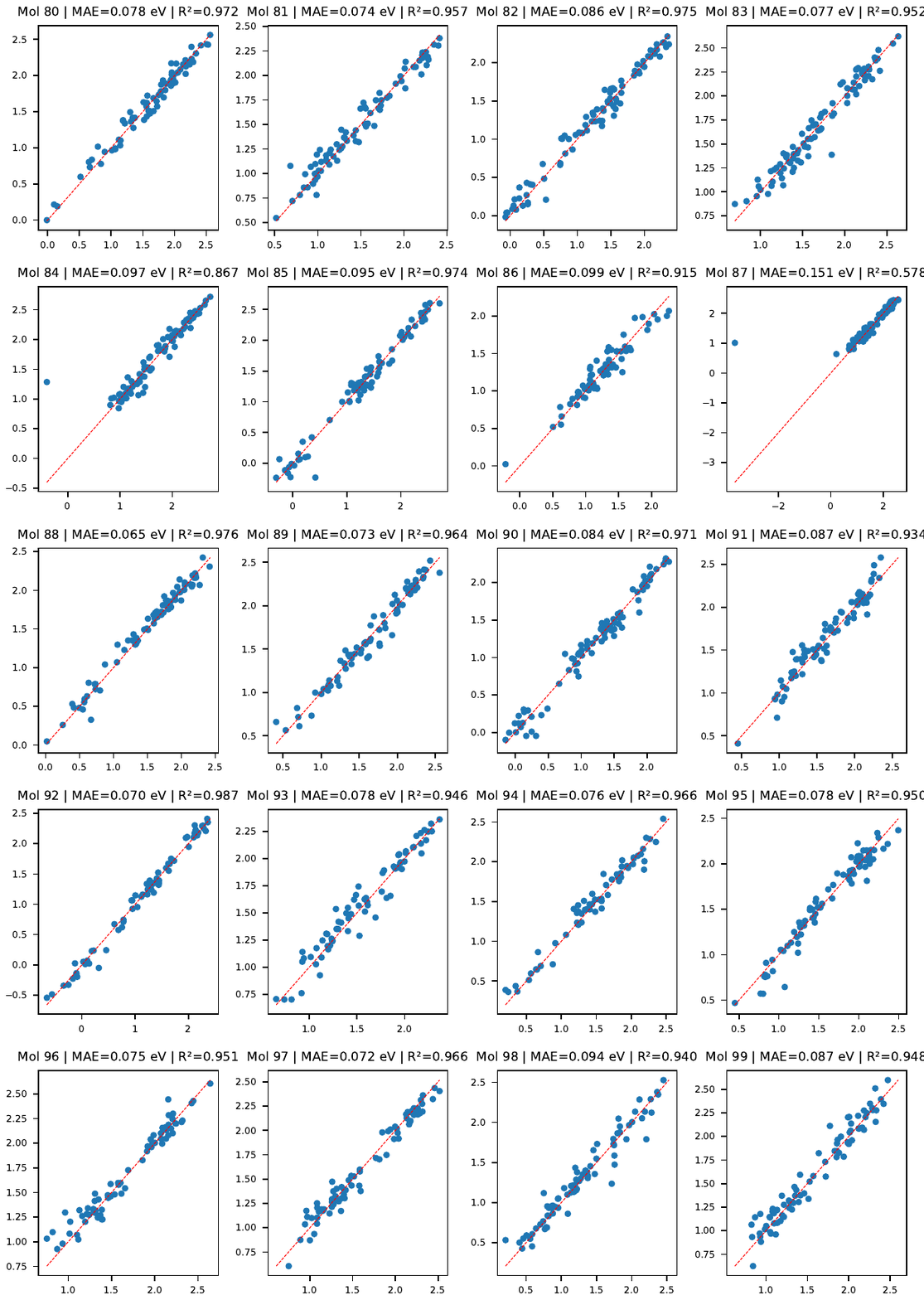}
\end{figure*}

\begin{figure*}[p]
    \centering
    \includegraphics[page=6,width=\textwidth]{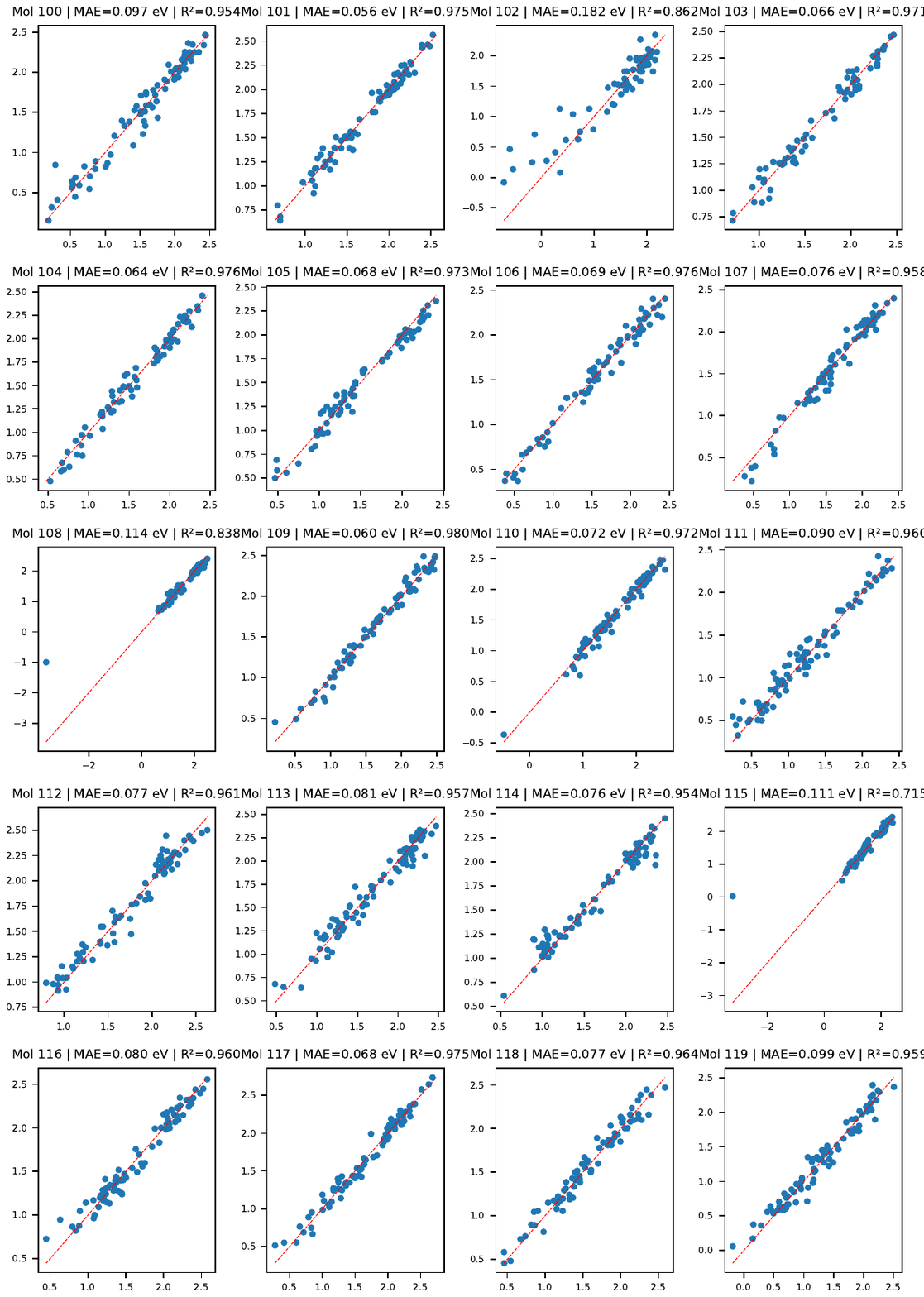}
\end{figure*}

\begin{figure*}[p]
    \centering
    \includegraphics[page=7,width=\textwidth]{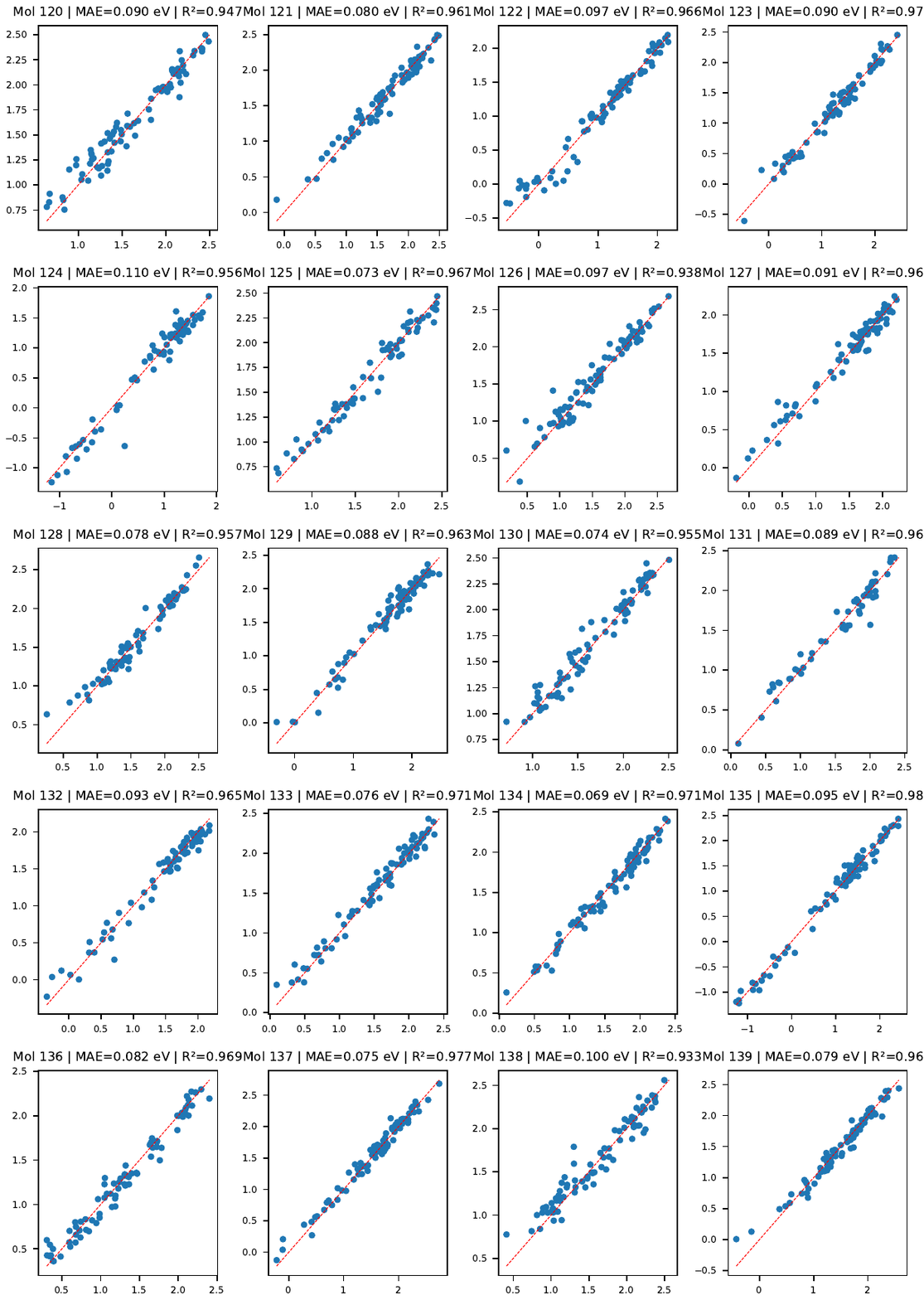}
\end{figure*}

\begin{figure*}[p]
    \centering
    \includegraphics[page=8,width=\textwidth]{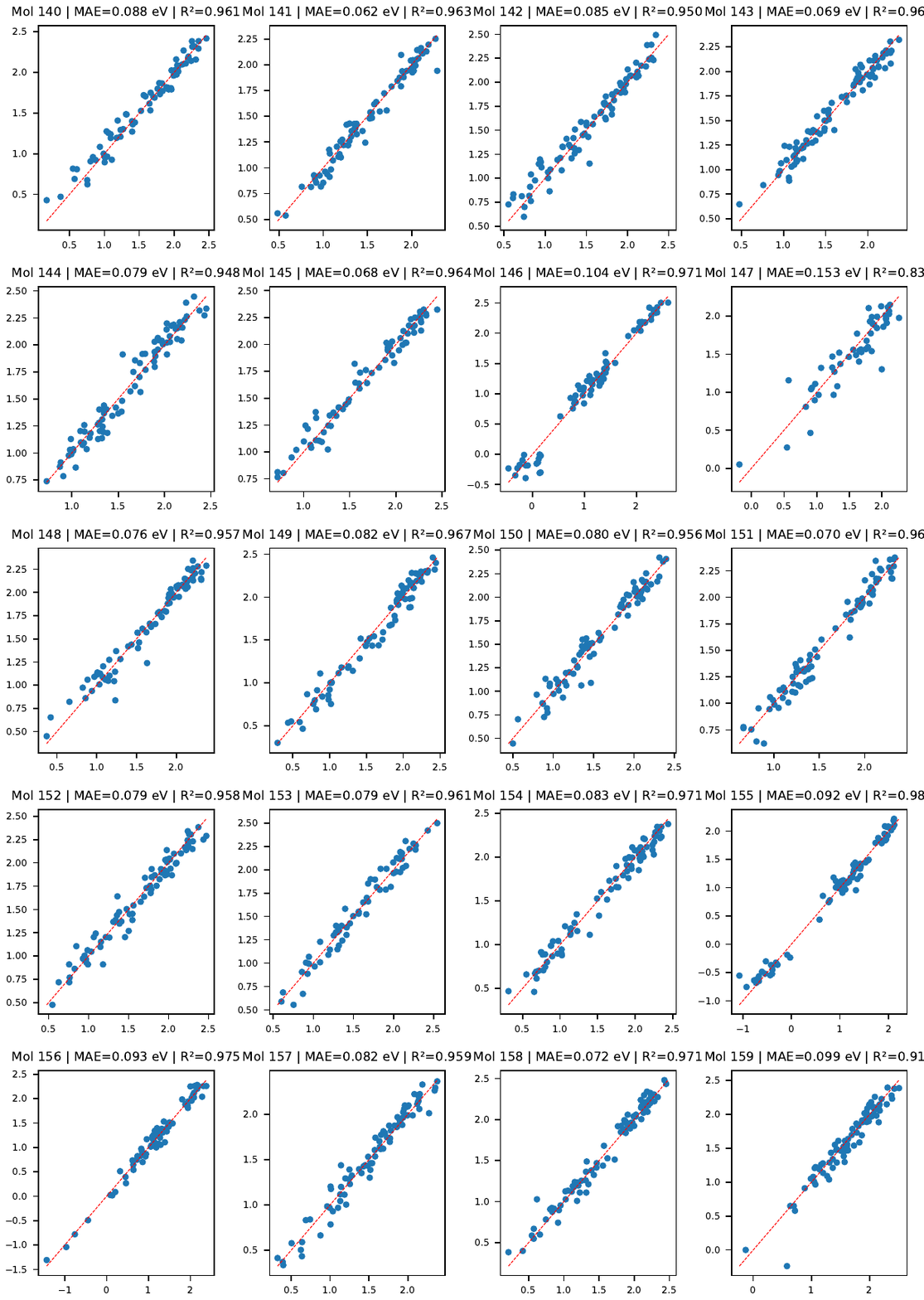}
\end{figure*}

\begin{figure*}[p]
    \centering
    \includegraphics[page=9,width=\textwidth]{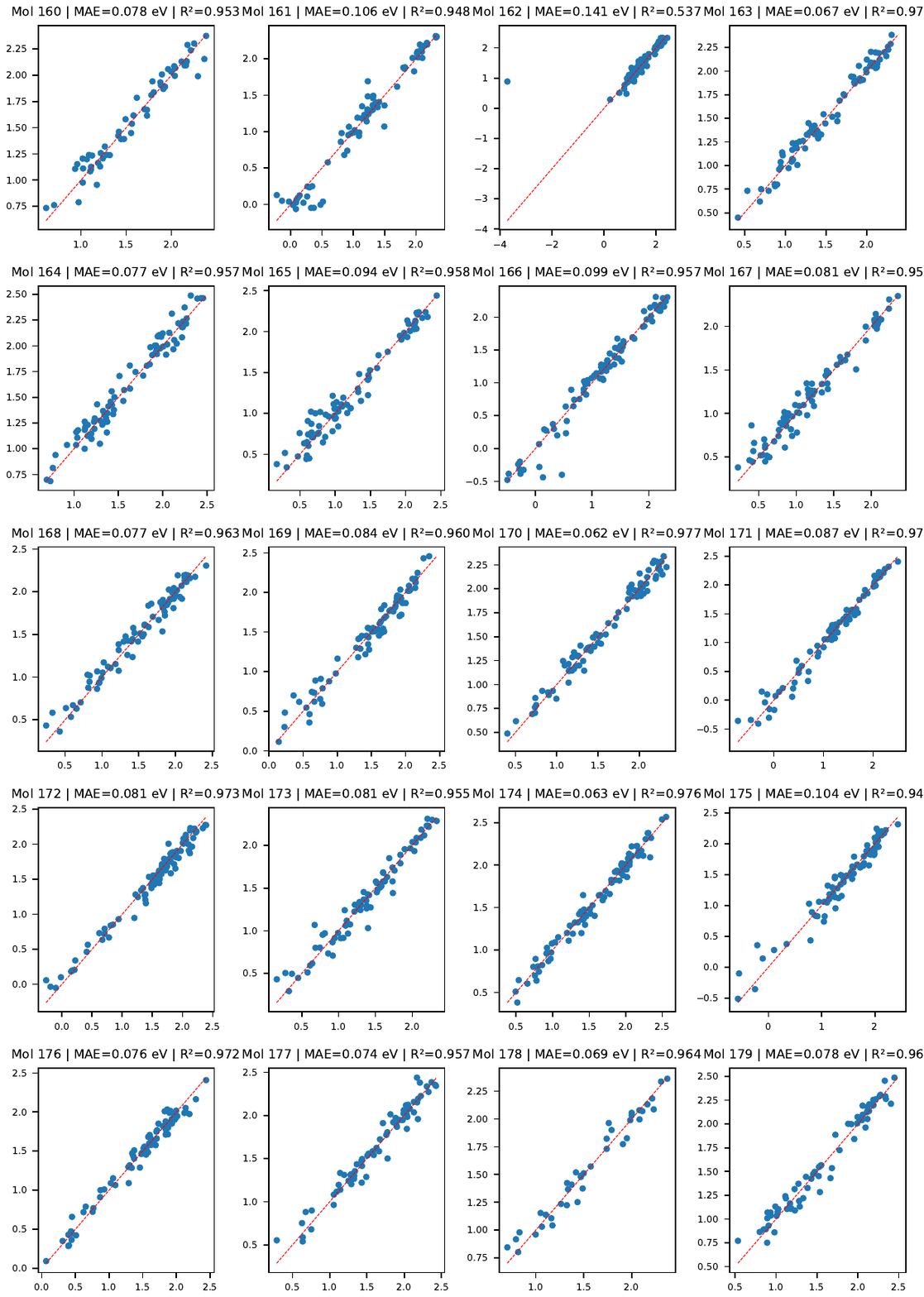}
\end{figure*}

\begin{figure*}[p]
    \centering
    \includegraphics[page=10,width=\textwidth]{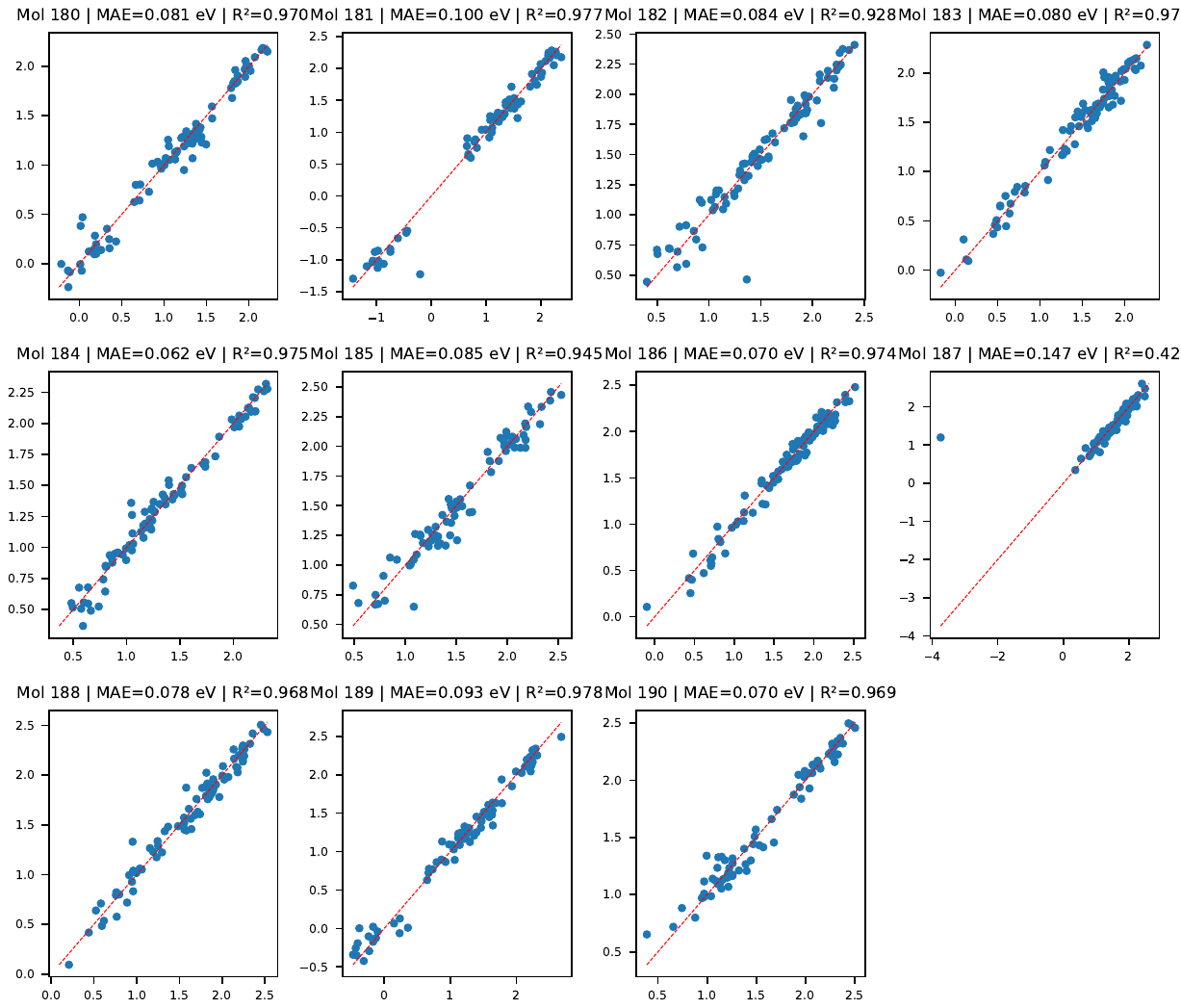}
\end{figure*}
\FloatBarrier

\section{Initial Conditions Selection and SHARC Input Files}\label{SI-SH}
The initial conditions for the dynamics simulations were created by sampling points from xtb trajectories (GFN2-xTB\cite{bannwarth2019gfn2}) at 300 K, where each 10th step was included. The total dynamics length was 100,000 steps. For all reported steps, the energies of the S$_0$ and S$_1$ were predicted using the trained X-MACE model, the oscillator strengths were predicted using the oscillator-strength MACE model. The initial conditions were then created by transcribing the trajectory data into the SHARC\cite{mai2018nonadiabatic,mai2025sharc} initconds format, and using a variation of SHARCs excite.py script, which reads in the existing initconds file, and uses the predicted oscillator strengths and energies to simulate delta-pulse excitation. The excitation window was set to 1.6 to 3.3 eV to cover the visible spectrum and excite all possible variants . For every excitation processed through this script the random seed used was 1234. The trajectories were then setup through the SHARC internal script.

The nonadiabatic molecular dynamics simulations were performed using the SHARC-md package. Initial geometries and velocities were read from external files (geom, veloc), written from the initconds.excited file. Two singlet states were included in the dynamics and treated as active ($S_0$ and S$_1$), enabling nonadiabatic population transfer between them. The initial electronic coefficients were generated automatically with a random seed generated by the setup script from the initial random seed 1234 to ensure reproducibility. The total charge was set to $-1$. The total simulation time was 7000 fs with a nuclear time step of 0.5 fs and 25 electronic substeps per nuclear step, ensuring stable integration of the electronic degrees of freedom.

Dynamics were carried out in the diagonal (adiabatic) representation, with nonadiabatic couplings treated using the curvature-driven approach. Gradient corrections were included, and the kinetic energy adjustment upon hopping was performed along the direction of the velocity vector. Frustrated hops were not reflected. Decoherence effects were treated using the energy-based decoherence correction scheme with a decoherence parameter of 0.1. Surface hops were performed using the standard SHARC algorithm. Gradients were computed for all states at each step. 

\section{Lifetime Fitting}
The rate constant for the population transfer from S$_1$ to S$_0$ were fitted using scripts implemented in SHARC using the Runge-Kutta 5th order algorithm to fit the kinetic parameter.\cite{kutta1901beitrag,runge1895numerische} The fitting of the model was performed on trajectories, which were cut after being in the S$_0$ for 10 fs.  The initial guess used for the fitting was 1000 fs.

\begin{figure}
    \centering
    \includegraphics[width=0.8\linewidth]{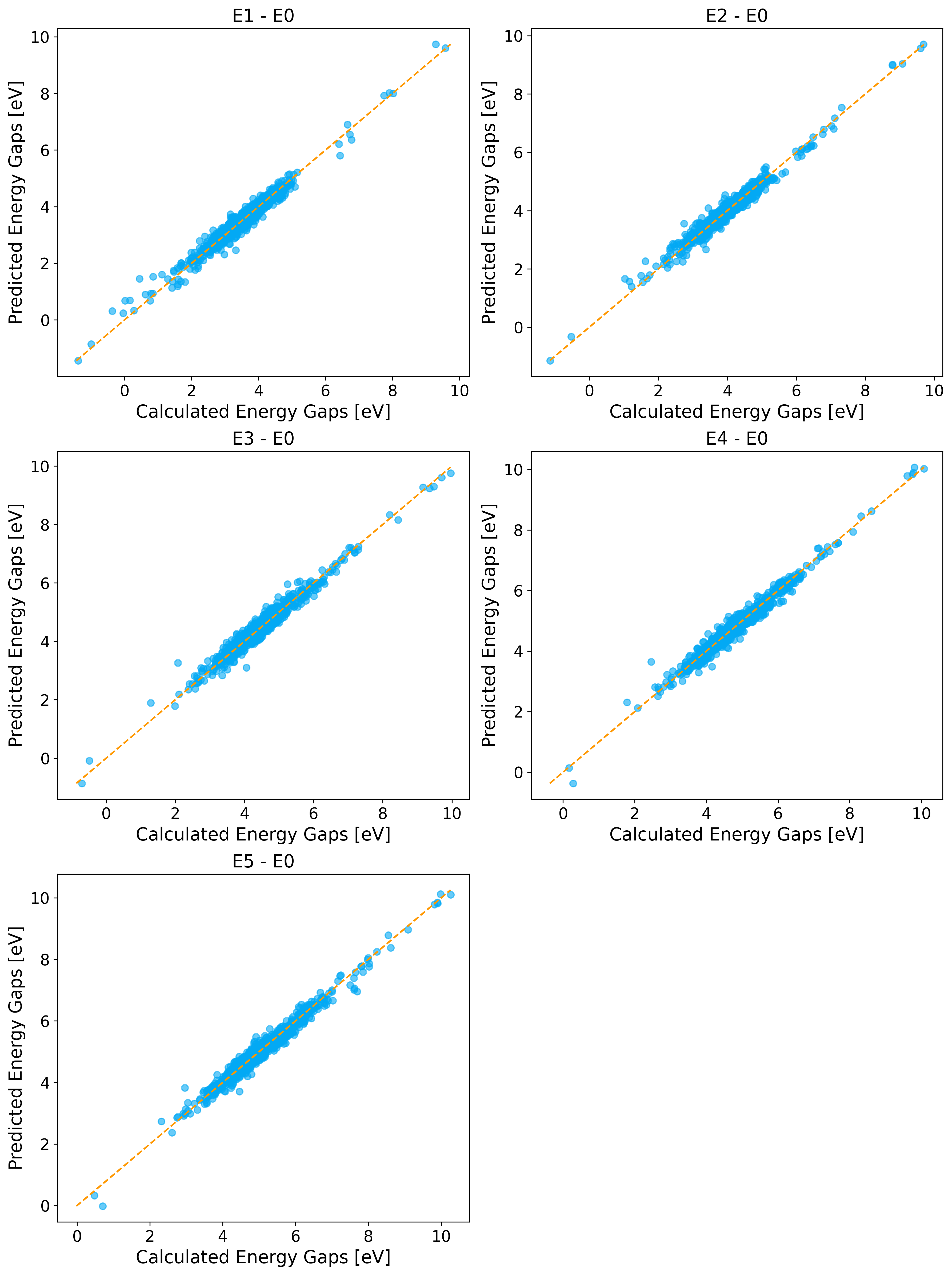}
    \caption{Parity plots comparing calculated and predicted vertical excitation energy gaps (in eV) for the excited states E$_1$--E$_5$, each referenced to the ground state E$_0$. The dashed orange lines indicate perfect agreement between calculated and predicted values ($y=x$).}
    \label{fig:chromophores-scatterplots-energies}
\end{figure}

\begin{figure}
    \centering
    \includegraphics[width=0.5\linewidth]{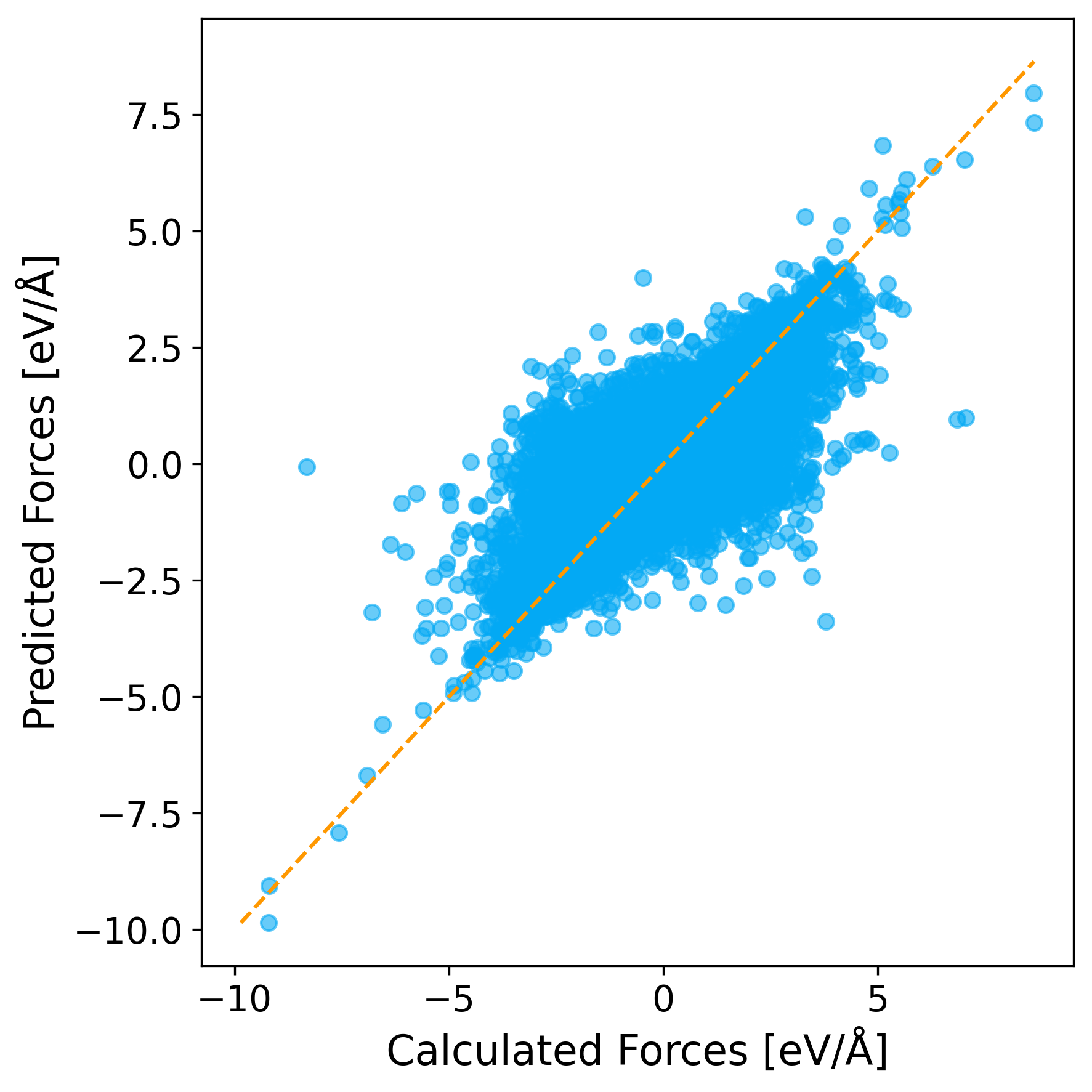}
    \caption{Parity plot of predicted versus calculated forces (in eV/\AA) for all atoms across the E$_1$--E$_5$ states. The dashed orange line represents perfect agreement ($y=x$).}    \label{fig:chromophores-scatterplots-forces}
\end{figure}

\section{References}

\bibliography{refs.bib} 

@article{martin2004origin,
  title={Origin, nature, and fate of the fluorescent state of the green fluorescent protein chromophore at the CASPT2//CASSCF resolution},
  author={Martin, Mar{\'\i}a Elena and Negri, Fabrizia and Olivucci, Massimo},
  journal={J. Am. Chem. Soc.},
  volume={126},
  number={17},
  pages={5452--5464},
  year={2004},
  publisher={ACS Publications}
}

@article{bey2017efficient,
  title={Efficient CNS targeting in adult mice by intrathecal infusion of single-stranded AAV9-GFP for gene therapy of neurological disorders},
  author={Bey, Karim and Ciron, Carine and Dubreil, Laurence and Deniaud, Johan and Ledevin, Mireille and Cristini, Joseph and Blouin, V and Aubourg, P and Colle, Marie-Anne},
  journal={Gene Ther.},
  volume={24},
  number={5},
  pages={325--332},
  year={2017},
  publisher={Nature Publishing Group}
}

@article{nielsen2001absorption,
  title={Absorption spectrum of the green fluorescent protein chromophore anion in vacuo},
  author={Nielsen, S Br{\o}ndsted and Lapierre, A and Andersen, Jens Ulrik and Pedersen, UV and Tomita, S and Andersen, LH},
  journal={Phys. Rev. Lett.},
  volume={87},
  number={22},
  pages={228102},
  year={2001},
  publisher={APS}
}

@article{luo2002subthalamic,
  title={Subthalamic GAD gene therapy in a Parkinson's disease rat model},
  author={Luo, Jia and Kaplitt, Michael G and Fitzsimons, Helen L and Zuzga, David S and Liu, Yuhong and Oshinsky, Michael L and During, Matthew J},
  journal={Science},
  volume={298},
  number={5592},
  pages={425--429},
  year={2002},
  publisher={American Association for the Advancement of Science}
}

@article{herndon2002stochastic,
  title={Stochastic and genetic factors influence tissue-specific decline in ageing C. elegans},
  author={Herndon, Laura A and Schmeissner, Peter J and Dudaronek, Justyna M and Brown, Paula A and Listner, Kristin M and Sakano, Yuko and Paupard, Marie C and Hall, David H and Driscoll, Monica},
  journal={Nature},
  volume={419},
  number={6909},
  pages={808--814},
  year={2002},
  publisher={Nature Publishing Group UK London}
}

@article{ho2024atomic,
  title={Atomic cluster expansion without self-interaction},
  author={Ho, Cheuk Hin and Gutleb, Timon S and Ortner, Christoph},
  journal={J. Comput. Phys.},
  volume={515},
  pages={113271},
  year={2024},
  publisher={Elsevier}
}

@article{yang2019analyzing,
  title={Analyzing learned molecular representations for property prediction},
  author={Yang, Kevin and Swanson, Kyle and Jin, Wengong and Coley, Connor and Eiden, Philipp and Gao, Hua and Guzman-Perez, Angel and Hopper, Timothy and Kelley, Brian and Mathea, Miriam and others},
  journal={J. Chem. Inf. Model.},
  volume={59},
  number={8},
  pages={3370--3388},
  year={2019},
  publisher={ACS Publications}
}

@article{dusson2022atomic,
  title={Atomic cluster expansion: Completeness, efficiency and stability},
  author={Dusson, Genevieve and Bachmayr, Markus and Cs{\'a}nyi, G{\'a}bor and Drautz, Ralf and Etter, Simon and van Der Oord, Cas and Ortner, Christoph},
  journal={J. Comput. Phys.},
  volume={454},
  pages={110946},
  year={2022},
  publisher={Elsevier}
}

@incollection{gilmer2020message,
  title={Message passing neural networks},
  author={Gilmer, Justin and Schoenholz, Samuel S and Riley, Patrick F and Vinyals, Oriol and Dahl, George E},
  booktitle={Machine learning meets quantum physics},
  pages={199--214},
  year={2020},
  publisher={Springer}
}

@inproceedings{gilmer2017neural,
  title={Neural message passing for quantum chemistry},
  author={Gilmer, Justin and Schoenholz, Samuel S and Riley, Patrick F and Vinyals, Oriol and Dahl, George E},
  booktitle={ICML},
  pages={1263--1272},
  year={2017},
  organization={Pmlr}
}

@article{mai2018nonadiabatic,
  title={Nonadiabatic dynamics: The SHARC approach},
  author={Mai, Sebastian and Marquetand, Philipp and Gonz{\'a}lez, Leticia},
  journal={Wiley Interdiscip. Rev. Comput. Mol. Sci.},
  volume={8},
  number={6},
  pages={e1370},
  year={2018},
  publisher={Wiley Online Library}
}

@article{gastegger2017machine,
  title={Machine learning molecular dynamics for the simulation of infrared spectra},
  author={Gastegger, Michael and Behler, J{\"o}rg and Marquetand, Philipp},
  journal={Chem. Sci.},
  volume={8},
  number={10},
  pages={6924--6935},
  year={2017},
  publisher={Royal Society of Chemistry}
}

@article{svendsen2017origin,
  title={Origin of the intrinsic fluorescence of the green fluorescent protein},
  author={Svendsen, Annette and Kiefer, Hjalte V and Pedersen, Henrik B and Bochenkova, Anastasia V and Andersen, Lars H},
  journal={J. Am. Chem. Soc.},
  volume={139},
  number={25},
  pages={8766--8771},
  year={2017},
  publisher={ACS Publications}
}

@inproceedings{batatia2022mace,
  title     = {MACE: Higher Order Equivariant Message Passing Neural Networks for Fast and Accurate Force Fields},
  author    = {Batatia, Ilyes and Kov{\'a}cs, D{\'a}vid P{\'e}ter and Simm, Gregor N. C. and Ortner, Christoph and Cs{\'a}nyi, G{\'a}bor},
  booktitle = {NeurIPS},
  year      = {2022},
  url       = {https://papers.nips.cc/paper_files/paper/2022/file/4a36c3c51af11ed9f34615b81edb5bbc-Paper-Conference.pdf}
}

@article{axelrod2022natcomm,
  title   = {Excited state non-adiabatic dynamics of large photoswitchable molecules using a chemically transferable machine learning potential},
  author  = {Axelrod, Simon and Wang, Xi and others},
  journal = {Nat. Commun.},
  year    = {2022},
  volume  = {13},
  number  = {1},
  pages   = {3310},
  doi     = {10.1038/s41467-022-30999-w}
}

@article{roos1980casscf,
  title   = {The complete active space SCF method in a finite basis set: A quasi-degenerate two-configuration example},
  author  = {Roos, Bj{\"o}rn O. and Taylor, Peter R. and Sigbahn, Per E. M.},
  journal = {Chem. Phys.},
  year    = {1980},
  volume  = {48},
  number  = {2},
  pages   = {157--173},
  doi     = {10.1016/0301-0104(80)80145-0}
}

@article{neese2025software,
  title={Software update: the ORCA program system—version 6.0},
  author={Neese, Frank},
  journal={Wiley Interdiscip. Rev. Comput. Mol. Sci.},
  volume={15},
  number={2},
  pages={e70019},
  year={2025},
  publisher={Wiley Online Library}
}

@article{kovacs2025mace,
  title={Mace-off: Short-range transferable machine learning force fields for organic molecules},
  author={Kov{\'a}cs, D{\'a}vid P{\'e}ter and Moore, J Harry and Browning, Nicholas J and Batatia, Ilyes and Horton, Joshua T and Pu, Yixuan and Kapil, Venkat and Witt, William C and Magdau, Ioan-Bogdan and Cole, Daniel J and others},
  journal={J. Am. Chem. Soc.},
  volume={147},
  number={21},
  pages={17598--17611},
  year={2025},
  publisher={ACS Publications}
}

@article{shimomura2022discovery,
  title={Discovery of green fluorescent protein, GFP. Nobel Lecture, 2008},
  author={Shimomura, O},
  journal={http://nobelprize.org/prizes/chemistry/2008/press-release/},
  year={2022}
}

@article{mai2025sharc,
    author = {Mai, S. and Bachmair, B. and Gagliardi, L. and Gallmetzer, H.-G. and Grünewald, L. and Hennefarth, M. R. and Høyer, N. M. and Korsaye, F. A. and Mausenberger, S. and Oppel, M. and Pitesa, T. and Polonius, S. and Gil, E.s. and Shu, Y. and Singer, N. K. and Tiefenbacher, M. X. and Truhlar, D. G. and Vörös, D. and Zhang, L. and González, L.},
    title ={SHARC4.0: Surface Hopping Including Arbitrary Couplings – Program Package for Non-Adiabatic Dynamics},
    journal ={https://sharc-md.org},
    year = {2025}
}

@article{list2022internal,
  title={Internal conversion of the anionic GFP chromophore: in and out of the I-twisted S 1/S 0 conical intersection seam},
  author={List, Nanna H and Jones, Chey M and Mart{\'\i}nez, Todd J},
  journal={Chem. Sci.},
  volume={13},
  number={2},
  pages={373--385},
  year={2022},
  publisher={Royal Society of Chemistry}
}

@article{bannwarth2019gfn2,
  title={GFN2-xTB—An accurate and broadly parametrized self-consistent tight-binding quantum chemical method with multipole electrostatics and density-dependent dispersion contributions},
  author={Bannwarth, Christoph and Ehlert, Sebastian and Grimme, Stefan},
  journal={J. Chem. Theory Comput.},
  volume={15},
  number={3},
  pages={1652--1671},
  year={2019},
  publisher={ACS Publications}
}

@article{landrum2013rdkit,
  title={Rdkit documentation},
  author={Landrum, Greg and others},
  journal={Release},
  volume={1},
  number={1-79},
  pages={4},
  year={2013}
}

@article{joung2020experimental,
  title={Experimental database of optical properties of organic compounds},
  author={Joung, Joonyoung F and Han, Minhi and Jeong, Minseok and Park, Sungnam},
  journal={Sci. Data},
  volume={7},
  number={1},
  pages={295},
  year={2020},
  publisher={Nature Publishing Group UK London}
}

@article{dreuw2015algebraic,
  title={The algebraic diagrammatic construction scheme for the polarization propagator for the calculation of excited states},
  author={Dreuw, Andreas and Wormit, Michael},
  journal={Wiley Interdiscip. Rev. Comput. Mol. Sci.},
  volume={5},
  number={1},
  pages={82--95},
  year={2015},
  publisher={Wiley Online Library}
}

@article{westermayr2022deep,
  title={Deep learning study of tyrosine reveals that roaming can lead to photodamage},
  author={Westermayr, Julia and Gastegger, Michael and V{\"o}r{\"o}s, D{\'o}ra and Panzenboeck, Lisa and Joerg, Florian and Gonz{\'a}lez, Leticia and Marquetand, Philipp},
  journal={Nat. Chem.},
  volume={14},
  number={8},
  pages={914--919},
  year={2022},
  publisher={Nature Publishing Group UK London}
}

@article{runge1895numerische,
  title={{\"U}ber die numerische Aufl{\"o}sung von Differentialgleichungen},
  author={Runge, Carl},
  journal={Mathematische Annalen},
  volume={46},
  number={2},
  pages={167--178},
  year={1895},
  publisher={Springer}
}

@book{kutta1901beitrag,
  title={Beitrag zur n{\"a}herungsweisen Integration totaler Differentialgleichungen},
  author={Kutta, Wilhelm},
  year={1901},
  publisher={Teubner}
}

@article{list2024chemical,
  title={Chemical control of excited-state reactivity of the anionic green fluorescent protein chromophore},
  author={List, Nanna H and Jones, Chey M and Mart{\'\i}nez, Todd J},
  journal={Commun. Chem.},
  volume={7},
  number={1},
  pages={25},
  year={2024},
  publisher={Nature Publishing Group UK London}
}

@article{meech2009excited,
  title={Excited state reactions in fluorescent proteins},
  author={Meech, Stephen R},
  journal={Chemical Society Reviews},
  volume={38},
  number={10},
  pages={2922--2934},
  year={2009},
  publisher={Royal Society of Chemistry}
}

@article{nienhaus2016chromophore,
  title={Chromophore photophysics and dynamics in fluorescent proteins of the GFP family},
  author={Nienhaus, Karin and Nienhaus, G Ulrich},
  journal={J. Phys. Condens. Matter},
  volume={28},
  number={44},
  pages={443001},
  year={2016},
  publisher={IOP Publishing}
}

@article{conyard2011chemically,
  title={Chemically modulating the photophysics of the GFP chromophore},
  author={Conyard, Jamie and Kondo, Minako and Heisler, Ismael A and Jones, Garth and Baldridge, Anthony and Tolbert, Laren M and Solntsev, Kyril M and Meech, Stephen R},
  journal={J. Phys. Chem. B},
  volume={115},
  number={6},
  pages={1571--1577},
  year={2011},
  publisher={ACS Publications}
}

@article{ashworth2023alkylated,
  title={Alkylated green fluorescent protein chromophores: dynamics in the gas phase and in aqueous solution},
  author={Ashworth, Eleanor K and Kao, Min-Hsien and Anst{\"o}ter, Cate S and Riesco-Llach, Gerard and Blancafort, Llu{\'\i}s and Solntsev, Kyril M and Meech, Stephen R and Verlet, Jan RR and Bull, James N},
  journal={Phys. Chem. Chem. Phys.},
  volume={25},
  number={35},
  pages={23626--23636},
  year={2023},
  publisher={Royal Society of Chemistry}
}

@article{shu2022nonadiabatic,
  title={Nonadiabatic dynamics algorithms with only potential energies and gradients: Curvature-driven coherent switching with decay of mixing and curvature-driven trajectory surface hopping},
  author={Shu, Yinan and Zhang, Linyao and Chen, Xiye and Sun, Shaozeng and Huang, Yudong and Truhlar, Donald G},
  journal={J. Chem. Theory Comput.},
  volume={18},
  number={3},
  pages={1320--1328},
  year={2022},
  publisher={ACS Publications}
}

\end{document}